\documentclass[useAMS,usenatbib]{mn2e}
\usepackage{graphicx}
\usepackage{amssymb}
\usepackage{amsmath}

\title[H$ \alpha $ Kinematics of NGC 864]
  {H$ \alpha $ Kinematics of S$ ^{4} $G spiral galaxies-I. NGC 864}
\author[S. Erroz-Ferrer et al.]
{Santiago Erroz-Ferrer,$^{1,2}$\thanks{Email: serroz@iac.es} Johan H. Knapen,$^{1,2}$ Joan Font,$^{1,2}$ John E. Beckman,$^{1,2}$  
\newauthor Jes\'us Falc\'on-Barroso,$^{1,2}$ Jos\'e Ram\'on S\'anchez-Gallego,$^{1,2}$ E. Athanassoula,$^{3}$
\newauthor Albert Bosma,$^{3}$ Dimitri A. Gadotti,$^{4}$  Juan Carlos Mu\~noz-Mateos,$^{5}$ Kartik Sheth,$^{5,6,7}$ 
\newauthor Ronald J. Buta,$^{8}$  S\'ebastien Comer\'on,$^{9}$ Armando Gil de Paz,$^{10}$ Joannah L. Hinz,$^{11}$  
\newauthor Luis C. Ho,$^{12}$ Taehyun Kim,$^{4,5,13}$ Jarkko Laine,$^{14}$ Eija Laurikainen,$^{14,15}$
\newauthor Barry F. Madore,$^{12}$ Kar\'in Men\'endez-Delmestre,$^{16}$ Trisha Mizusawa,$^{5,6,7}$
\newauthor Michael W. Regan,$^{17}$ Heikki Salo $^{14}$ and Mark Seibert$^{12}$\\
$^{1}${Instituto de Astrof\'isica de Canarias, V\'ia L\'actea s/n 38205 La Laguna, Spain}\\
$^{2}${Departamento de Astrof\'isica, Universidad de La Laguna, 38206 La Laguna, Spain}\\
$^{3}${Aix Marseille Universit\'e CNRS, LAM (Laboratoire d'Astrophysique de Marseille) UMR 7326, 13388, Marseille, France}\\
$^{4}${European Southern Observatory, Casilla 19001, Santiago 19, Chile}\\
$^{5}${National Radio Astronomy Observatory / NAASC, 520 Edgemont Road, Charlottesville, VA 22903, USA}\\
$^{6}${Spitzer Science Center, 1200 East California Boulevard, Pasadena, CA 91125, USA}\\
$^{7}${California Institute of Technology, 1200 East California Boulevard, Pasadena, CA 91125, USA}\\
$^{8}${Department of Physics and Astronomy, University of Alabama, Box 870324, Tuscaloosa, AL 35487, USA} \\
$^{9}${Korea Astronomy and Space Science Institute, 61-1 Hwaam-dong, Yuseong-gu, Daejeon 305-348, Republic of Korea}\\
$^{10}${Departamento de Astrof\'isica, Universidad Complutense de Madrid, Madrid 28040, Spain}\\
$^{11}${Steward Observatory, University of Arizona, 933 N. Cherry Ave, Tucson, AZ 85721, USA}\\
$^{12}${The Observatories of the Carnegie Institution for Science, 813 Santa Barbara Street, Pasadena, CA 91101, USA}\\
$^{13}${Astronomy Program, Department of Physics and Astronomy, Seoul National University, Seoul 151-742, Republic of Korea}\\
$^{14}${Division of Astronomy, Department of Physical Sciences, University of Oulu, Oulu, FIN-90014, Finland}\\
$^{15}${Finnish Centre of Astronomy with ESO (FINCA), University of Turku, V\"ais\"al\"antie 20, FI-21500, Piikki\"o, Finland}\\
$^{16}${Universidade Federal do Rio de Janeiro, Observat\'orio do Valongo, Ladeira Pedro Ant\^onio, 43, CEP 20080-090, Rio de Janeiro, Brazil}\\
$^{17}${Space Telescope Science Institute, 3700 San Martin Drive, Baltimore, MD 21218, USA}}

\date{Accepted 2012 July 23. Received 2012 July 9; in original form 2012 June 5}

\pagerange{\pageref{firstpage}--\pageref{lastpage}} \pubyear{2012}

\begin{document}

%\label{firstpage}

\maketitle

\begin{abstract}
We present a  study of the kinematics of the isolated spiral galaxy NGC 864, using H$ \alpha $ Fabry-Perot data obtained with the Galaxy H$ \alpha $ Fabry-Perot System (GH$ \alpha $FaS) instrument at the William Herschel Telescope in La Palma, complemented with images at 3.6 $ \mu  $m, in the \textit{R} band and in H$ \alpha $ filter, and integral-field spectroscopic data. The resulting data cubes and velocity maps allow the study of the kinematics of the galaxy, including in-depth investigations of the rotation curve, velocity moment maps, velocity residual maps, gradient maps and position-velocity diagrams. We find asymmetries in the velocity field in the bar zone, caused by non-circular motions, probably in response to the potential of the bar. We also find a flat-profile bar, in agreement with the strong bar, with the grand design spiral pattern, and with the gap between the ends of the bar and the start of the spiral arms. We quantify the rate of massive star formation, which is concentrated in the two spiral arms.
\end{abstract}


\section{Introduction}

Galaxies are the basic building blocks of the Universe and their formation and evolution are of great interest in current astrophysical research. Their dynamics and morphology are the result of both externally driven (e.g. galaxy mergers) and secular evolution (e.g. bar or spiral pattern driven). Disentangling the different early evolutionary tracks is complicated by redshift, distance and dust absorption. Alternatively one can explore the fossil record of evolution through detailed observations of the end product, nearby galaxies.

One of the most important drivers of this internal secular evolution is the flow of gas into the central regions. This is the result of angular momentum loss in shocks induced by non-axisymmetric potentials, as generated by a bar, interactions, minor mergers or even by minor deviations from axisymmetry such as ovals, lenses, lopsidedness or spiral arms (e.g. \citealt{Schwarz1984}; \citealt{Shlosman1989,Shlosman1990}; \citealt{Knapen1995}; \citealt{Rix1995};   \citealt{Kormendy2004}; \citealt{Comeron2010}). Around two-thirds of local galaxies exhibit at least one bar (\citealt{RC3}; \citealt{Sellwood1993}; \citealt{Moles1995}; \citealt{Ho1997}; \citealt{Mulchaey1997}, \citealt{Hunt1999}; \citealt{Knapen2000}; \citealt{Eskridge2000}; \citealt{Laine2002}; \citealt{Laurikainen2004a}; \citealt{Menendez-Delmestre2007}; \citealt{Marinova2007}; \citealt{Sheth2008}; \citealt{Laurikainen2009}), so potential drivers of secular evolution are common. Low-mass discs acquired their bars later than more massive ones (\citealt{Sheth2008}), implying that the bar-driven secular evolution is expected to be different depending on each galaxy's mass.

To study the influence of the past evolution of a galaxy on the observed morphology, kinematic information is essential. The study of velocity fields in spirals has mostly been done using the 21 -cm HI line, primarily because this emission can be traced far out, often three or four times beyond the visible disc. \citet{Rots1975}, \citet{vanderHulst1979}, \citet{Bosma1981},  \citet{Gottesman1982} and others demonstrated the power of this technique in deriving the total mass distribution of disc galaxies. Since the beam sizes associated with 21 cm observations were typically at least as large as the disc scale length of the galaxy, little small-scale structure in the velocity field was detected in this manner. The HI Nearby Galaxy Survey (THINGS; \citealt{Walter2008}) presents some of the highest spatial resolution ($\sim6''$) 21-cm HI observations of nearby galaxies to date, using the Very Large Array (VLA) of the National Radio Astronomy Observatory (NRAO).

The best candidate technique to yield two-dimensional kinematical maps across a whole galaxy with good arcsecond spatial resolution is H$\alpha$. Due to the cosmic abundance of hydrogen, H$\alpha$ is often the brightest emission line in the visible wavelength range. In spiral galaxies, this line traces primarily the ionized gas in HII regions around young massive stars. Traditional long-slit spectra with optical emission lines are valuable for deducing the rotation curve of a galaxy (e.g., \citealt{Rubin1980}, \citealt{Rubin1982a}, \citealt{Rubin1985}, \citealt{Amram1992}, \citealt{Amram1994}, or \citealt{Persic1995}), and the rotation curves can be extended to very large radii using H$ \alpha $ (e.g. \citealt{Christlein2008}). H$\alpha$ observations using Fabry-Perot (FP) instruments have been used for some 40 years now (\citealt{Tully1974}; \citealt{Deharveng1975}, \citealt{Dubout1976} or \citealt{deVaucouleurs1980}), and have since been employed to create high signal-to-noise ratio (S/N) rotation curves for many spiral galaxies (e.g., \citealt{Marcelin1985}; \citealt{Bonnarel1988}; \citealt{Pence1990}; \citealt{Corradi1991}; \citealt{Cecil1992}; \citealt{Sicotte1996}; \citealt{Ryder1998}; \citealt{Jimenez-Vicente1999}; \citealt{Garrido2002}, \citealt{Epinat2008}, and many more). 

Using the Infrared Array Camera (IRAC; \citealt{Fazio2004}) operating at 3.6 and 4.5 $ \mu $m, the \textit{Spitzer} Survey of Stellar Structure in Galaxies (S$^4$G; \citealt{Sheth2010}) targets over 2300 galaxies. The sample is volume-, magnitude- and size-limited (\textit{d}$<$40 Mpc, \textit{$m_b$}$<$15.5, \textit{$D_{25}$}$>$1 arcmin). The cornerstone of the survey is the quantitative analysis of photometric parameters, enabling a variety of studies on secular evolution, outer disc and halo formation, galaxy morphology etc. The new images, much deeper than traditional ground-based near-IR observations, will allow a comprehensive and definitive study of galaxy structure not only as a function of stellar mass, but also as a function of environment, vital to test cosmological simulations predicting the mass properties of present day galaxies.

We have designed an observing programme to obtain H$\alpha$ FP kinematic data sets of over 40 spiral galaxies that are included in the S$ ^{4} $G sample. The observation of NGC 864 is part of this wider programme using the instrument Galaxy H$ \alpha $ Fabry-Perot System (GH$ \alpha $FaS) at the William Herschel Telescope (WHT) in La Palma. NGC 864 is used to illustrate the data and methods, and discuss the kinds of results the main survey will give.

%%%%%%%%%%%%%%%%%%%%%%%%%%%%%%%%%%%%%%%%%%%%%%%%%%

\section{Target selection}

The galaxies in our FP programme had to satisfy the following requirements: first of all, the galaxy should fit well in the GH$\alpha$FaS field of view (FOV) of 3.4 $\times$ 3.4 arcmin. Therefore, galaxies with diameters between 2.7 and 3.4 arcmin were selected. The declination should be higher than -10$\degr$ so that the altitude in the sky can be enough at the time of the observation in La Palma, the Hubble type should be later than Sab (T=2; RC3, \citealt{RC3}) to ensure H$\alpha$ emission, and only galaxies with inclinations between 0$\degr$ and 70$\degr$ were selected. 

These criteria yielded a sample of 108 S$^4$G galaxies, from which the observed sample was selected by requiring a spread in morphological type, galaxy mass (from absolute magnitude; ten galaxies in each of four bins), bar presence, visibility on the sky and finally, the availability of ancillary data (interferometric CO, HI, ultraviolet, \textit{Spitzer} mid-IR or \textit{Herschel}). The galaxies observed in our first observing run were NGC 428, NGC 864, NGC 918, NGC 1073 and NGC 7479. This paper will concentrate on the galaxy NGC 864.

NGC 864 is a spiral galaxy, well isolated from nearby, similar-sized companions [Catalog of Isolated Galaxies (CIG); \citealt{Karachentseva1973}], although it is a radial-velocity confirmed member of a small group (\citealt{Fouque1992}), including UGC 1670 ($v=$1601 $km/s$; $\Delta$M = 3.12 mag; $ \sim $ 84 arcmin or $ \sim $5.7 Mpc away from NGC 864) and UGC 1803 ($v=$ 1624 $km/s$; $\Delta$M = 3.0 mag; $ \sim $89 arcmin or $ \sim $6.1 Mpc away from NGC 864); all these values taken from the NASA/IPAC Extragalactic Database (NED). NGC 864 is classified as S\underline{A}B(rs)c by The de Vaucouleurs Atlas of Galaxies (\citealt{Buta2007}). \citet{Buta2005} found a maximum relative bar torque $ Q_{b} $=0.321, implying that the bar is fairly strong in spite of its apparent weakness in blue light. The main characteristics of the galaxy are presented in Table \ref{props}.

\begin{table}

\caption{Galaxy global properties. (1) The centre position was fixed to the position of the optical centre from \citet{Leon2003}. (2) The systemic velocity is listed as the central velocity of the HI spectrum measured at the 20\% level, from \citet{Espada2005}. (3) Distance calculated after applying the Virgo, GA and Shapley corrections, with H$_0$= 73 $\pm$ 5 km s$^{-1}$ Mpc$^{-1}$, from NASA/IPAC Extragalactic Database (NED). (4) Morphological type from \textit{The de Vaucouleurs Atlas of Galaxies}, (\citealt{Buta2007}). (5) \textit{B} magnitude from \textit{The Third Reference Catalogue of Bright Galaxies} (RC3; \citealt{RC3}). (6) Bar and Spiral Arm torques, from \citet{Buta2005}. (7) Disc inclination and position angle (PA) of the kinematic major axis, from the RC3. (8)  Disc inclination and PA obtained from the 25.5 mag/arcsec$ ^{2} $ isophote with the S$ ^{4} $G 3.6 $ \mu $m image, calculations and uncertainties explained and to be published in Mu\~noz-Mateos et al. (2012, in prep). (9) Disc inclination and PA from our H$ \alpha $ FP data (Section 4.2.1).  (10) Luminosity and Star Formation Rate (SFR) of the whole galaxy measured from our H$ _{\alpha} $ images (Section 4.3).}
 \label{props}
\centering
\begin{tabular}{|c|c|c|}
\hline

Name  & NGC 864   & Comments  \\
\hline 
  $\alpha_{J2000.0}$  &  $2^{h}15^{m}27.6^{s}$  & (1) \\ 
  $\delta_{J2000.0}$  &  +06$\degr00'09.1''$  &  (1)\\ 
 $v_{\rm sys}$ ($\rm km\,s^{-1}$)  & 1561.6   & (2) \\ 
 \textit{D} (Mpc)  &  20.9 $\pm$ 1.5  & (3) \\ 
  Type  &  S\underline{A}B(rs)c  &  (4)\\ 
  $m_{B}  $ (mag)  &  11.62 $ \pm $ 0.21   &  (5) \\ 
 \textit{Q$ _{b} $} &  0.321  &  (6)\\ 
  Q$ _{s} $ &  0.134  &  (6)\\ 
\textit{i$ _{RC3} $ }&  43\degr   &  (7) \\ 
     \textit{i$ _{3.6\mu m} $}   & 44.6\degr   & (8)  \\
    \textit{i$ _{H\alpha} $}  &  28.2\degr $ \pm $ 20.4\degr  & (9)  \\
 PA$ _{RC3} $  &  20\degr  &  (7) \\
   PA$ _{3.6\mu m} $  &  22.0\degr  & (8)  \\ 
      PA$ _{H\alpha} $  &  25.3 \degr $ \pm $ 6.7\degr  &  (9) \\
       $ L ({\rm H\alpha}) $ (erg s$ ^{-1} $) & $(4.0 \pm 1.6 )\cdot10 ^{41} $  & (10)  \\  
    SFR ($M _{\odot} $ yr$ ^{-1} $)  & 2.19 $ \pm $ 0.88  & (10)  \\
    \hline 
\end{tabular}
\end{table} 

This galaxy has been observed and analysed by many other authors, whose work we summarise here. \citet{Martini2003} studied visible and near-IR \textit{Hubble Space Telescope (HST)} images and color maps of NGC 864, identifying bright star formation rings. One of these is a small circumnuclear star-forming ring with a radius of less than 100 pc, confirmed by the Atlas of Images of NUclear Rings (AINUR; \citealt{Comeron2010}). \citet{Epinat2008} described the analysis of H$\alpha$ FP cubes of 97 galaxies included in their GHASP survey of 207 spiral galaxies, presenting the rotation curve of every galaxy, including NGC 864. \citet{Buta2009} found three corotation radii in the galaxy: one lies within the bar, the next is located slightly beyond where the bar ends and the third is within the inner spiral arms.

\citet{Espada2005} provide the most extensive and significant study of NGC 864, via HI observations. They measure large-scale asymmetries in the two-dimensional kinematics, finding that the galaxy's profiles are symmetric in velocity but asymmetric in intensity. These asymmetries are found in the outer parts of the disc, beyond the extent of our new data.

\section{Observations and data reduction}

The data for NGC 864 have been taken at different epochs using three different instruments of the 4.2-m WHT on the Roque de Los Muchachos Observatory in La Palma: the GH$\alpha$FaS instrument, the Auxiliary port CAMera (ACAM) and the  Spectrographic Areal Unit for Research on Optical Nebulae (SAURON) integral field spectrograph. These data are complemented with a 3.6 $ \mu $m \textit{Spitzer} image from the S$ ^{4} $G survey.

\subsection{ACAM}

One of the motivations of this paper is to develop a method to flux-calibrate GH$\alpha$FaS Fabry-Perot data cubes. To do this, we have obtained narrow-band imaging of the galaxy nearly simultaneously with the GH$\alpha$FaS data. NGC 864 was observed on 2010 September 4 with the ACAM instrument, permanently mounted at a folded-Cassegrain focus of the WHT \citep{Benn2008}. The FOV in imaging mode is 8 arcmin, with a pixel size of 0.25 arcsec.

Two images were taken with ACAM: a 30 s exposure image using a Johnson filter \textit{R}, with central wavelength 6228 \AA{} and FWHM 1322 \AA{}; and an H$\alpha$ image, with an exposure time of 300 s, using a narrow filter having a central wavelength 6589 \AA{} and FWHM 15 \AA{}. The [NII]-6548 line falls into the narrow filter transmission range (at 6582 \AA{}), with a similar throughput to the H$ \alpha $ line (which lies in 6597 \AA{}). However, the contribution of the [NII] line is not important, as in the majority of star forming regions in galaxy discs the [NII] fraction is small or negligible (\citealt{James2005}).

\begin{figure}
\begin{center}
\includegraphics[width=70mm]{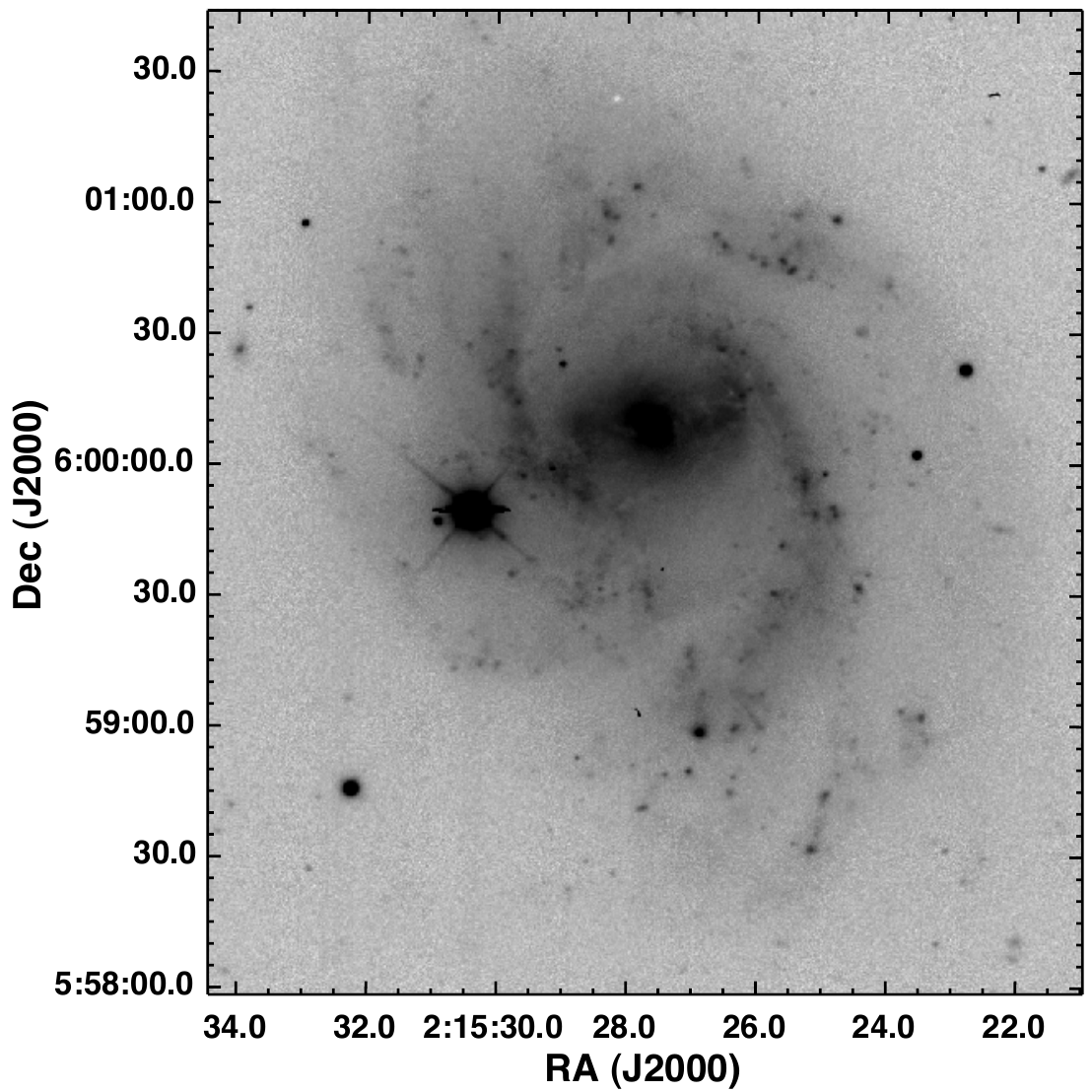}
\includegraphics[width=70mm]{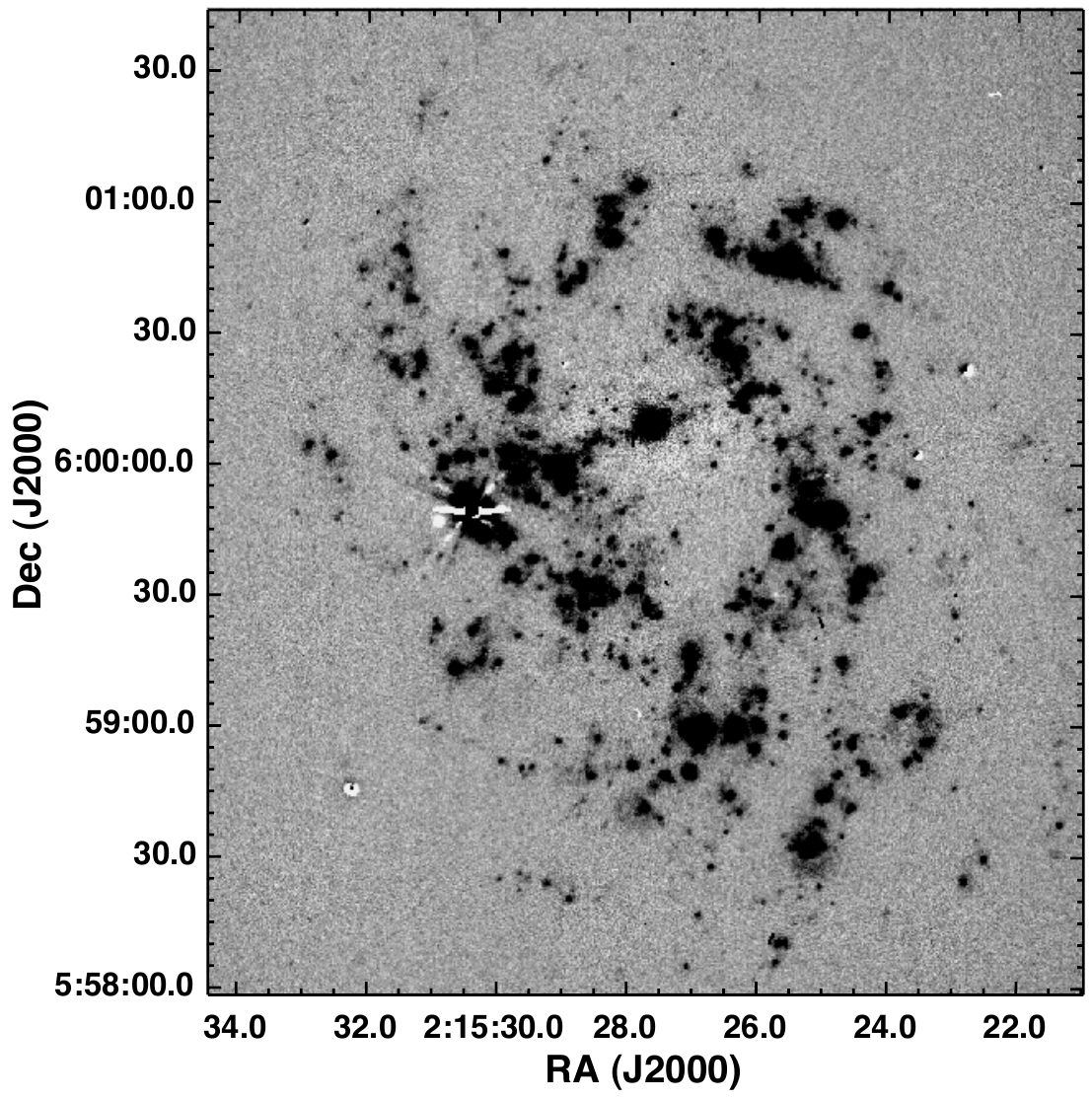}
 \caption{\textit{R}-band (\textit{top}) and continuum-subtracted H$\alpha$ (\textit{bottom}) images of NGC 864, taken with ACAM.}
\label{acam}
\end{center}
\end{figure}

The basic reduction of these images was performed mainly using IRAF. Firstly, bias and flat corrections were made. Then, the continuum was subtracted following the procedures outlined in \citet{Knapen2004} and the image was photometrically calibrated. To perform the continuum subtraction, the \textit{R}-band image was scaled by a factor of 0.0109 and subtracted from the H$\alpha$ image. The resulting images are shown in Figure \ref{acam}. 

\subsection{GH$ \alpha $FaS}

\subsubsection{Observations and data reduction}

The galaxy was observed for 4 hours on the same night as the ACAM images with GH$\alpha$FaS \citep{Hernandez2008}, situated in the Nasmyth focus of the WHT. GH$\alpha$FaS is a Fabry-Perot imaging interferometer that provides high spectral resolution H$\alpha$ images with a seeing-limited spatial resolution within a 3\farcm4 $\times$ 3\farcm4 FOV. The instrument is composed of a focal reducer, a filter wheel, a Fabry-Perot etalon, an image photon-counting system (IPCS) camera, and a calibration lamp (neon source). The etalon employed works at an interference order of 765 and a Free Spectral Range (FSR) of 391.9 km s$^{-1}$. Each observational cycle consists of stepping through the etalon for 10 seconds per channel. In the observation of NGC 864, 29 cycles of 48 channels/cycle were observed in total. The high spatial resolution mode was used, achieving a 8.167 km/s velocity resolution with 0\farcs2/pix scale in 1k$\times$1k pixel images.

As a proper de-rotator at the Nasmyth focus of the telescope is not provided for GH$ \alpha $FaS, the Fabry-Perot cubes need to be de-rotated before being further reduced. To do this, we have followed the work done by \citet{Blasco2010}. Two fixed points in the field are selected in every image (ideally stars), and used to de-rotate each plane. Lamp exposures are used to calibrate the observations in wavelength (see \citealt{Hernandez2008}). Due to the nature of the Fabry-Perot data, any two-dimensional transformation (translation or rotation) of the raw images has to be done three-dimensionally (including the third spectral dimension), and thus wavelength calibration and de-rotation must be performed simultaneously. The resulting de-rotated and  wavelength-calibrated data cube is then imported into GIPSY (Groningen Image Processing System; \citealt{vanderHulst1992}). The data are placed on an astrometrically correct spatial grid by comparing the positions of stars in the ACAM images, yielding a data cube of 48 planes of 1001 $\times$ 1001 pixels each. A very bright star nearby, approximately at RA $2^{h}15^{m}30^{s}$ and DEC $+05\degr59'49''$, was masked out with the GIPSY task \textit{BLOT}.

De-rotating blurs the image somewhat, and the FWHM after the de-rotation was determined to be 1\farcs56 along the RA-axis and 1\farcs4 along the DEC-axis (the worst seeing that night was around 1\farcs2, with stages of 0\farcs8 during the 4 hours of observation). The cube was smoothed to different resolutions with the GIPSY task \textit{SMOOTH}. We have worked at two different spatial resolutions. A first slight smoothing was done so that the image loses very little spatial resolution, hence taking advantage of the high spatial resolution of the instrument (high-resolution cube hereafter, with a FWHM of 1\farcs65 $ \times $ 1\farcs56). Then, a 3$'' \times 3''$ smoothing was performed in order to get continuous spatial information and to highlight low-emission regions, facilitating the computation of, for instance, the rotation curve (hereafter low-resolution cube). After the spatial smoothing, the continuum was subtracted using \textit{CONREM} of GIPSY, which estimates the continuum level from the line-free frames (the first three channels and the last seven in this case).

\begin{figure*}
\begin{center}
 %\includegraphics[width=168mm]{ghafasmapsheat.pdf}
%ONLINE VERSION:
 \includegraphics[width=168mm]{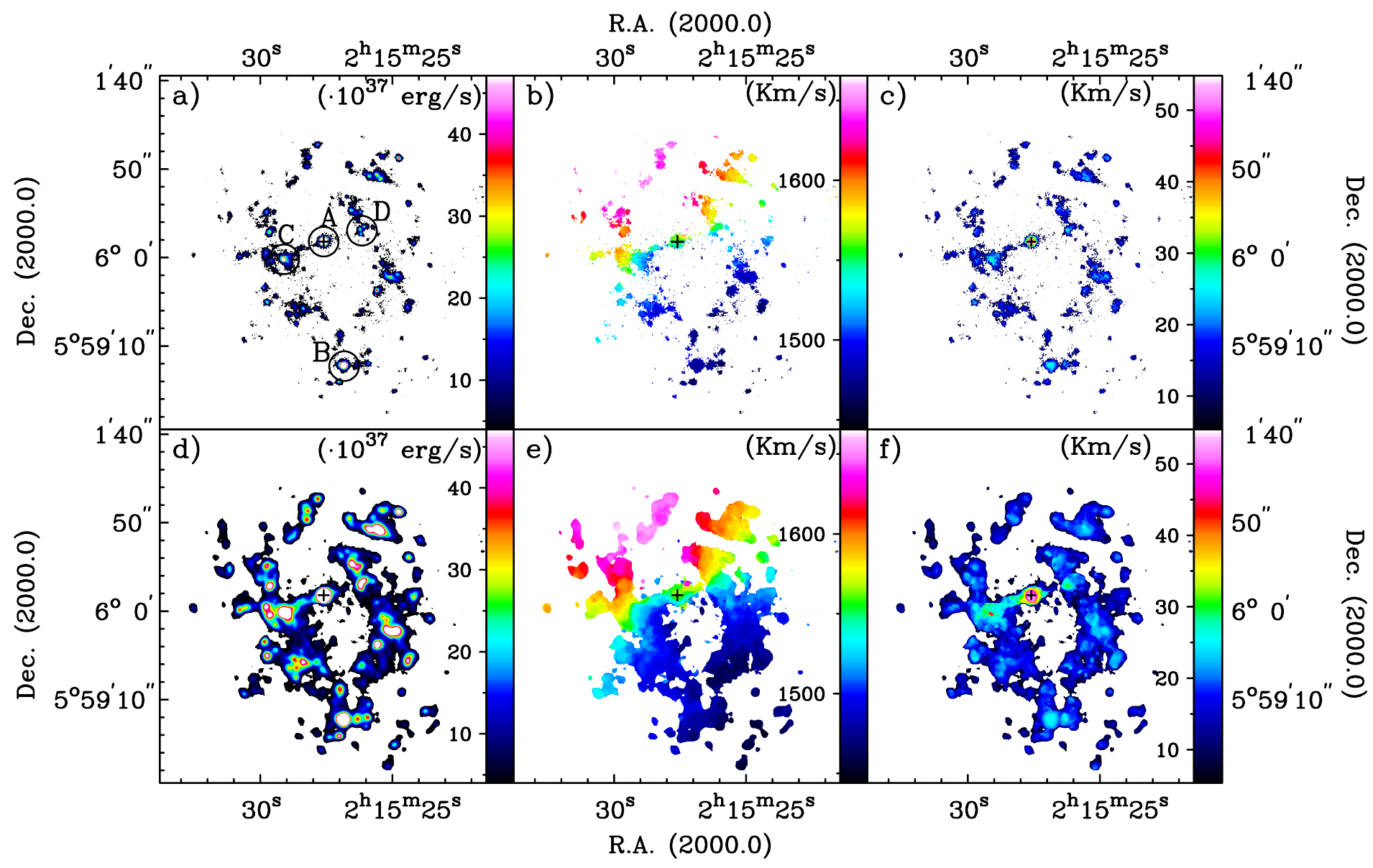}
\caption{\textit{a)} Intensity map for the high-resolution data. The centre is marked with A. B is the brightest HII region, located at the end of the western arm. C and D refer to other regions of massive star formation, found at the ends of the bar, and near the beginning of the arms. \textit{b)} Velocity map for the high-resolution data. \textit{c)} Dispersion velocity map for the high-resolution data. \textit{d)} Intensity map for the low-resolution data. \textit{e)} Velocity map for the low-resolution data. \textit{f)} Dispersion velocity map for the low-resolution data.  The centre of the galaxy is indicated with a cross, calibrated colours are labelled in the vertical colour bars.}
\label{ghafasmaps}
\end{center}
\end{figure*}

To create moment maps, the GIPSY task \textit{MOMENTS} was used. The moment maps start from the smoothed and continuum-subtracted cubes. We imposed two conditions: the signal has to be higher than 3$\sigma$, in at least three adjacent channels. The resulting moment maps are presented in Figure \ref{ghafasmaps}. There is a little leakage from the neighbouring [NII]-6548 line, that has to be taken into account when calibrating the flux. However, whenever a second peak is present, the task \textit{MOMENTS} does not take it into account because it only uses the main peak.

\subsubsection{Flux calibration}

\begin{figure}
\begin{center}
 \includegraphics[width=84mm]{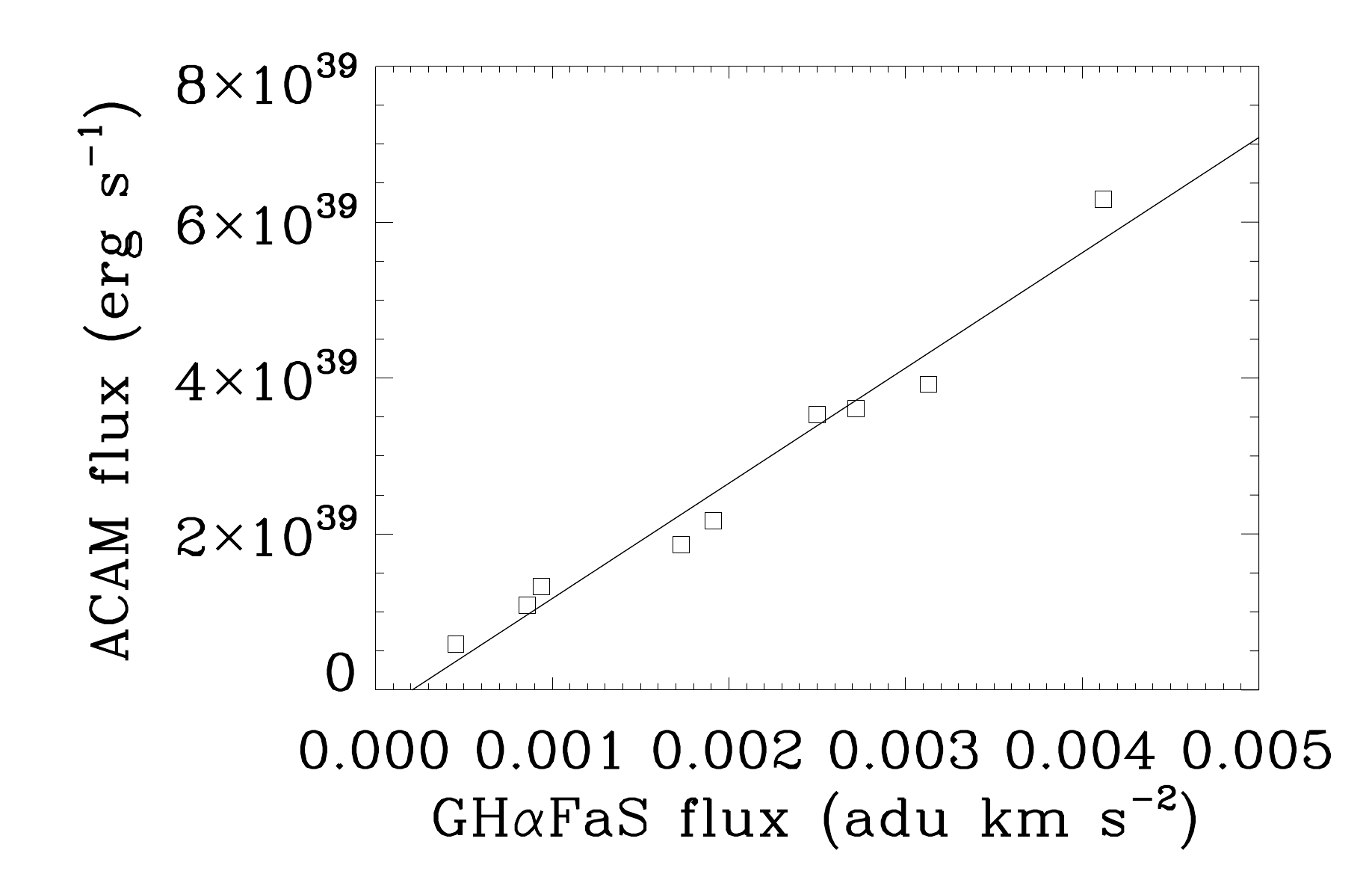}
\caption{Linear fit for the flux calibration. With squares, fluxes of some HII regions measured in ACAM H$ \alpha $ image presented versus the fluxes of the same regions measured in the GH$ \alpha $FaS H$ \alpha $ intensity map, whereas the straight line represents the line fitted to the data.}
\label{calib1}
\end{center}
\end{figure}

This paper presents the first attempt to flux-calibrate a GH$\alpha$FaS data cube. Previous observations with the instrument were used for kinematic studies and therefore did not need to be flux-calibrated. The continuum-subtracted and flux-calibrated ACAM H$\alpha$  image was used to perform the flux calibration of the GH$\alpha$FaS data. Fluxes from selected HII regions in both the GH$\alpha$FaS cube and the ACAM image were measured, and then compared. The transmission of the filter has to be taken into account, assuming a Gaussian profile. As explained before, the ACAM H$ \alpha $ image contains a little leakage coming from the [NII]-6548 line, and the FP also used a filter that allowed light from that [NII]-line to come in. It is thus necessary to correct for these leaks, and the correction applied is of 3\% following the procedures of \citet{James2005}. Figure \ref{calib1} presents the linear fit to data points corresponding to a set of HII regions from the GH$\alpha$FaS and ACAM data.

The uncertainty in the flux calibration is 10\% to 20\%, arising mainly from the use of galaxy distance in the calibration of the ACAM image, and to a lesser degree from the method described above.

\subsection{SAURON}

NGC 864 was also observed with the two-dimensional integral-field spectrograph SAURON (\citealt{Bacon2001}), and these data of NGC 864 were presented in \citet{Ganda2006}. SAURON delivers a spectral resolution of 4.2 \AA{} FWHM, spatial sampling of 0.96 arcsec and covers the narrow spectral range 4800-5380 \AA{} (1.1 \AA{} pixel$^{-1}$). This wavelength range includes a number of important stellar absorption lines (e.g. H$\beta$, Fe, Mgb) and potential emission lines as well (H$\beta$, [OIII], [NI]).   The field of view, of 33 $\times$ 41 arcsec$^2$, is much smaller than those of GH$ \alpha $FaS or ACAM, but the instrument delivers stellar as well as ionized gas kinematic information simultaneously.

 \citet{Ganda2006} presented the stellar and gas kinematics of 18 nearby late-type galaxies, including NGC 864. The data for this galaxy were taken in 2004 January 20-26. Four exposures of 1800 s each were obtained using the low spatial resolution mode of SAURON, with the FOV mentioned before.

\begin{figure}
\begin{center}
% \includegraphics[width=84mm]{SAURONMAPSheat.pdf}
%ONLINE VERSION:
 \includegraphics[width=84mm]{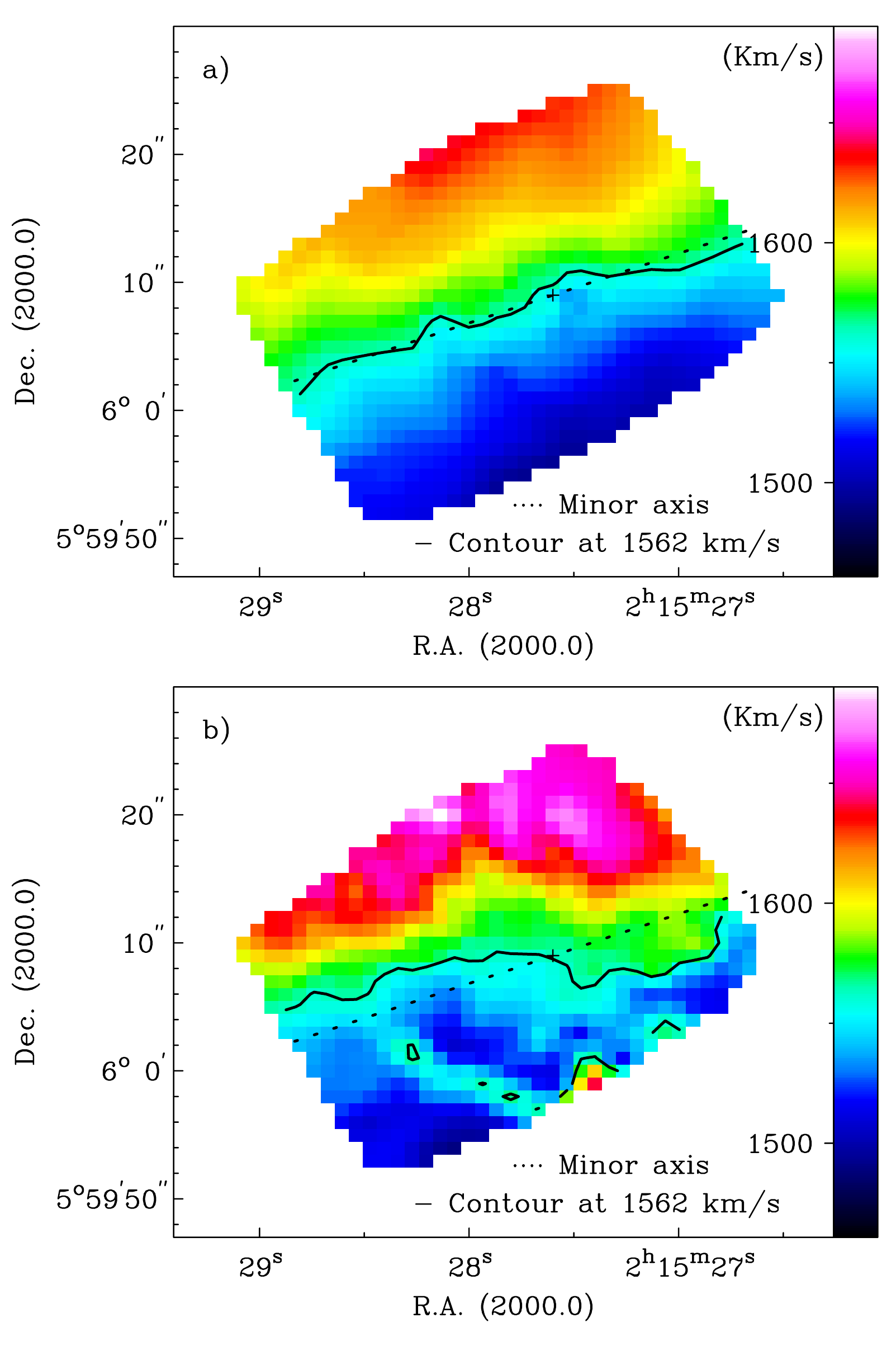}
\caption{\textit{a)} Velocity map of the stars in NGC 864, observed with SAURON. \textit{b)} Velocity map of the gas (H$\beta$) observed with SAURON. The kinematic minor axis (\textit{PA}=20º) is plotted with a dotted line, whereas the straight line shows the derived systemic velocity (see Section 4.2.1). The centre is indicated with a cross.}
\label{SAURONMAPS}
\end{center}
\end{figure}

\citet{Ganda2006} reduced the data using the dedicated software \textit{XSAURON}, including bias and dark subtraction, rotation of all the frames, extraction of the spectra using the fitted mask model, wavelength calibration, low-frequency flat-fielding and cosmic ray removal. Ater that, the data reduction consisted of homogenization of the spectral resolution over the FOV, sky subtraction, and finally flux calibration of the spectra, although the data were not necessarily collected under photometric conditions. In Figure \ref{SAURONMAPS}, the velocity maps for gas (H$\beta$) and stars (computed using the wavelength region that contains the H$\beta$, Fe5015 and Mgb absorption lines; see \citealt{Ganda2006} for more details) are presented.

\subsection{S$ ^{4} $G image}

In order to derive geometric galaxy parameters and to study specific galaxy features, a 3.6 $ \mu $m image from the S$ ^{4} $G survey has been used. This image was observed in the warm mission phase of Spitzer. As the galaxy diameter is lower than 3\farcm 3 and the Spitzer FOV is 5', the galaxy was mapped using a single pointing, following a standard cycling small dither pattern with 4 exposures of 30 seconds each in two separate Astronomical Observation Requests (AORs). The image has a pixel scale of 0\farcs75, with an angular resolution of 1\farcs7. The reduction pipeline is described in detail in \citet{Sheth2010}.

\begin{figure}
\centering
\includegraphics[width=70mm]{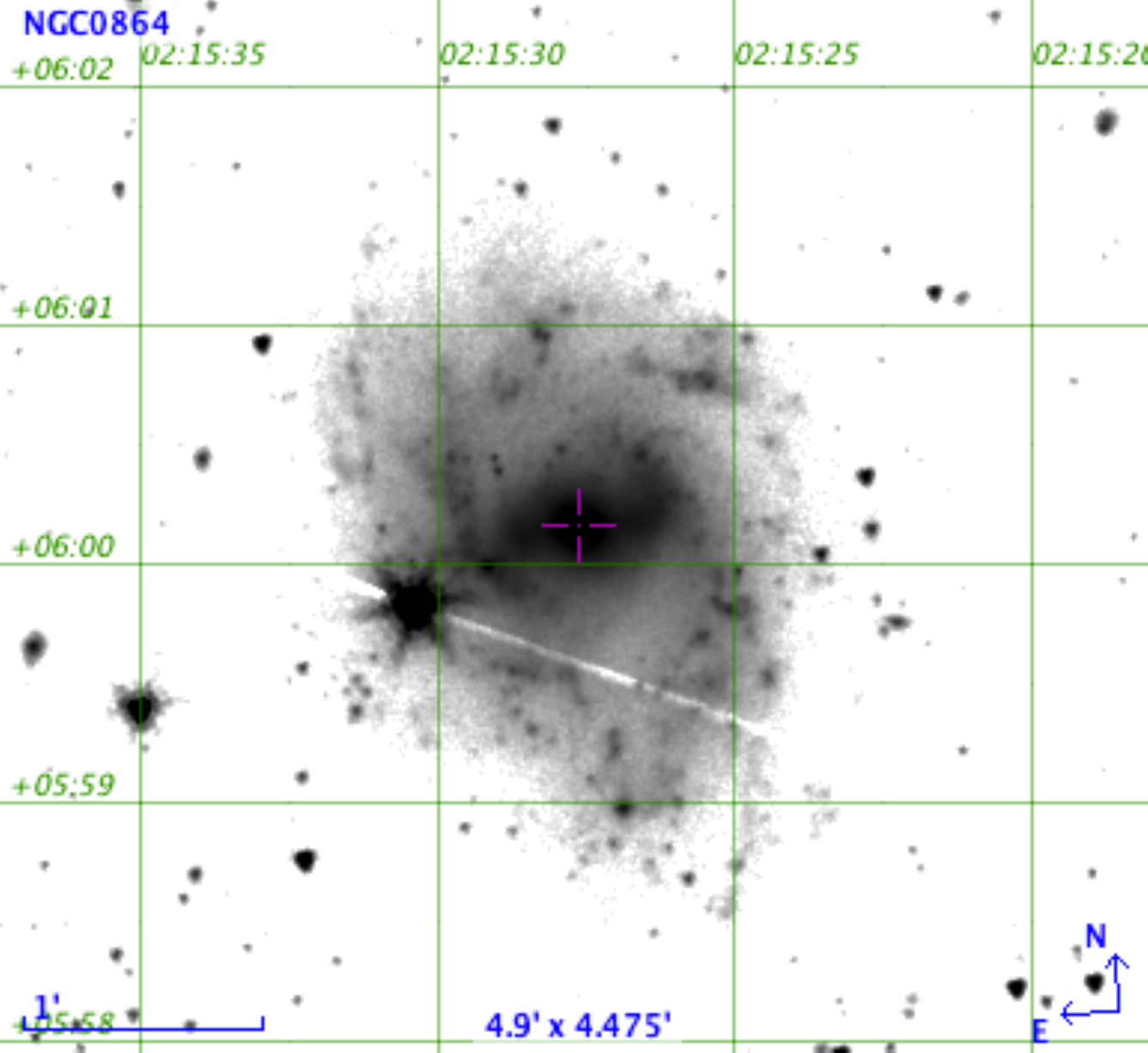}
\includegraphics[width=70mm]{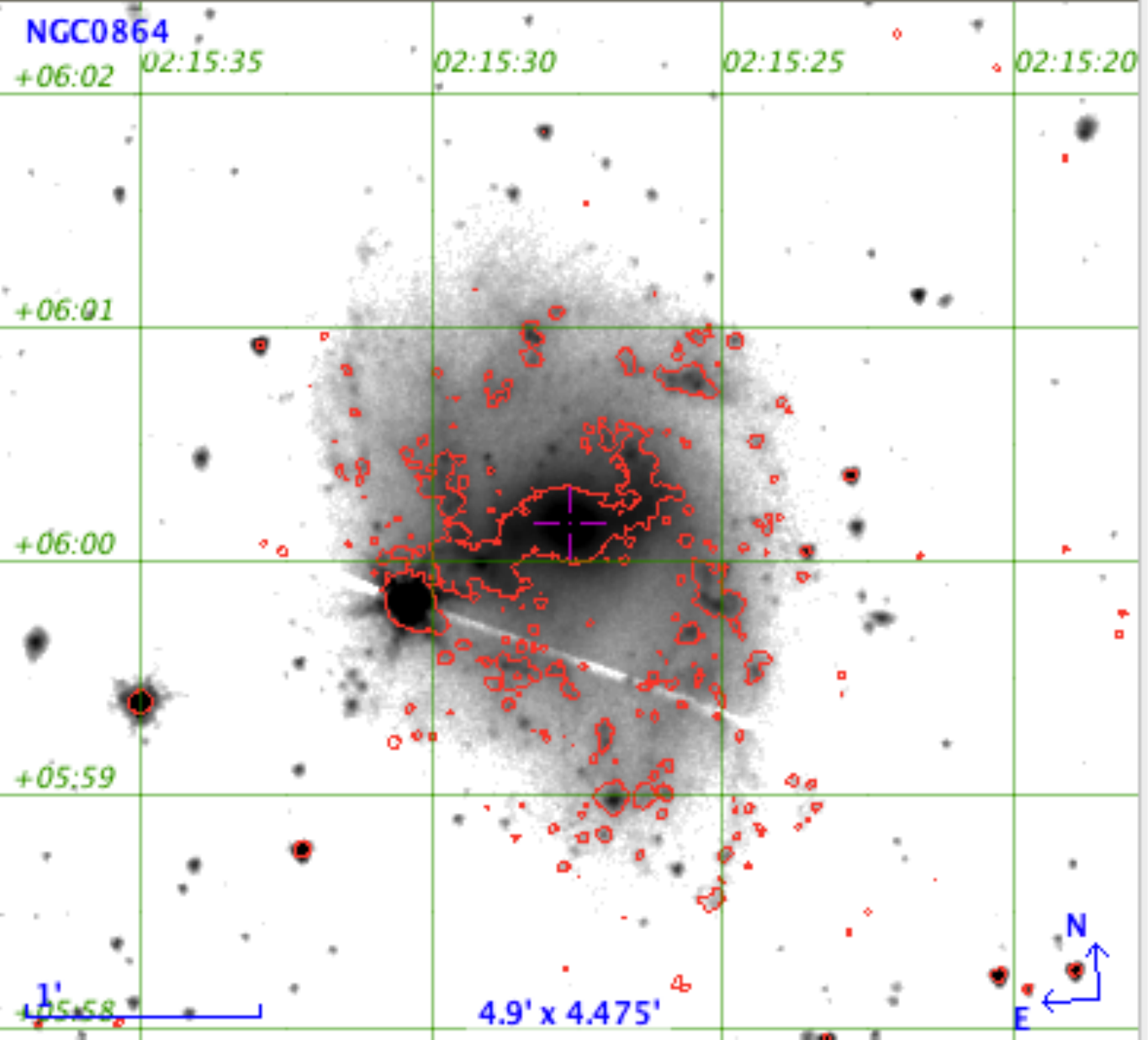}
\caption{\textit{Top}: 3.6 $ \mu $m S$ ^{4} $G image. \textit{Bottom}: the same with H$ \alpha $ contours at $1.1 \cdot 10^{36}$ erg/s, taken from the ACAM image.}
\label{s4g}
\end{figure}

%%%%%%%%%%%%%%%%%%%%%%%%%%%%%%%%%%%%%%%%%%%%%%%%%%

\section{Results}

\subsection{Morphology}

NGC 864 is a late-type barred spiral galaxy, with strong emission in the centre. In blue light, the type is S\underline{A}B(rs)c (\textit{The de Vaucouleurs Atlas of Galaxies}; \citealt{Buta2007}), but the mid-IR type is more like SA\underline{B}(r\underline{s})b or bc. The bar does look stronger in the S$  ^{4}$G image (Figure \ref{s4g}), but not as strong as in type SB. The bar is foreshortened, which undoubtedly contributes to its weak projected appearance.

The bar and some dust lanes can be distinguished in the ACAM \textit{R}-band image (Figure \ref{acam}). The bar presents bright regions of massive star formation at the ends, where two asymmetric arms arise. The arms do not, as is usually the case in galaxies, start at the end of the bar, but in fact overshoot that position, leading to a mismatch between the start of the spiral arms and the ends of the bar. We measure from the S$ ^{4} $G image that the eastern arm starts 0.89 $ \pm $  0.07 kpc away from the eastern end of the bar, while the western arm arises 0.96 $ \pm $ 0.07  kpc from the western end of the bar. This gap between the end of the bar and the start of the spiral arms may be the space between the inner resonances and corotation, as spiral arms may be excited only beyond corotation [like \citet{Elmegreen1985} found in some flat-profile bar spirals].  

The arms make less than a half-turn before disappearing. The outer spiral pattern is more flocculent than a two-armed spiral, but not as flocculent as e.g. NGC 2841 (\citealt{Block1996}). This was also noticed by \citet{Elmegreen1987}, as they classified NGC 864 to be a \textit{grand design} galaxy with \textit{Arm class} 5, i.e. two symmetric, short arms in the inner regions and irregular outer arms. 

The skewed appearance of the bar in NGC 864 seems to be a result of projection
effects. It is possible to distinguish both the outer flat part of the
bar (i.e. the part of the bar coplanar with the disc) and the inner
vertically thickened part of the bar (usually referred to as a
box/peanut 'bulge'). \citet{Athanassoula2006} presented \textit{N}-body
simulations of barred galaxies explaining how projection effects can
lead to the same configuration in the case of M31. In NGC 864, the
flat outer part of the bar has a position angle (PA) of about 90-100 degrees (counter clockwise from North),
whereas the inner part of the bar has a PA of around 70-80 degrees.

Alternatively, the skewed appearance of the bar might be real. If we inspect the bar starting from the galaxy
centre, it looks straight initially and then curls a little like an
open spiral, although the overshoot of the main arms could contribute
to this. In addition, the inner section looks like a
twisted oval, and not much like a box with a subtle X-pattern. In
the de Vaucouleurs Atlas of Galaxies \citep{Buta2007}, clearer
cases of low inclination examples of this are present.

We have performed a bulge/disc/bar 2D decomposition of the S$^4$G
3.6$\mu m$ image of NGC 864, using {\sc budda}
\citep{Souza2004,Gadotti2008}, in order to check whether the light
profile along the bar is exponential or rather flat.
\citet{Elmegreen1985} found that bars in galaxies with
morphological types later than Sc show exponential light profiles,
whereas bars in galaxies with earlier types show flat profiles. The
RC3 classification of NGC 864 is Sc, and thus it is clearly a
borderline case. The fit is complicated by the projection effects
mentioned above and the rather unusual start position of the spiral
arms. We find that the S\'ersic index of the bar is 0.24, meaning that the
bar profile is flat, and even somewhat brightening at the ends, although
this is likely an effect of an enhanced luminosity at the ends due to
a pair of bar ansae and the beginning of the spiral arms. Nevertheless, the
bar profile is certainly not exponential. To verify the
robustness of these results, several fits were performed with varying
initial parameters. Although the results of each of these fits show
variations for a few parameters, the bar S\'ersic index is always lower than 0.5, confirming a flat light profile, with high
confidence.

\citet{Elmegreen2011} present a study of spiral arm properties of 46 galaxies in the S$ ^{4} $G sample, which in effect forms a modern-day extension of the pioneering work of \citet{Elmegreen1985}. In their sample, all of the flocculent barred galaxies have exponential bars, while all of the grand design barred galaxies have flat bars. NGC 864 is a grand design galaxy with a flat profile and thus fits into this picture.

\citet{Athanassoula2002} showed that flat profiles are found in strong bars and exponential ones in weaker bars. This fits in with the results as presented by \citet{Combes1993}, who found from simulations that flat and exponential bars are linked to the shape of the rotation curve and halo concentration, finding that flat bars tend to occur in galaxies with sharply rising inner rotation curves (as we find in NGC 864 - see Section 4.2), while exponential bars tend to occur in galaxies with slowly rising rotation curves. \citet{Athanassoula2003} showed that this is due to the fact that sharply rising rotation curves are linked to a strong spheroidal component in the inner parts, and a considerable amount of spheroidal mass around the main resonances. This leads to significant angular momentum exchange within the galaxy, which in turn can lead to strong bars. This was further discussed by \citet{Athanassoula2009a}, who modeled the two bar types and compared them to observations made by \citet{Gadotti2007}.

NGC 864 is a clear case where a flat-profile bar is found in a two-armed spiral with a strong bar. The fact that there is a gap between the ends of the bar and the beginning of the spiral arms is also consistent with flat-profile bars (\citealt{Elmegreen1985}). The bar does indeed appear to be of the ansae type, which is very unusual for a late-type galaxy (\citealt{Martinez-Valpuesta2007}). This means that NGC 864 is an example of an early-type bar in a late-type galaxy. Its spiral structure, however, is not very grand design even in the mid-IR. Also, it is not an ansae type in the typical manner since it looks like an open spiral.

The moment 0 maps (left plots in Figure \ref{ghafasmaps}) and the ACAM H$ \alpha $ image (Figure \ref{acam}) highlight the regions with the most H$\alpha$ emission. We note two main emitting regions marked in Figure \ref{ghafasmaps}, panel \textit{a}: the centre (marked with A), and the brightest HII region located at the end of the western arm (letter B). The regions where the arms start and the bar ends have also been marked with C (east)  and D (west). Such morphology is common in barred galaxies.

\subsection{Kinematics}

We analysed moment maps at high and low resolution from the FP GH$\alpha$FaS observations, and gaseous and stellar moment maps are from the SAURON observations.

The patchiness of the H$\alpha$ emission maps does not allow us to analyse the maps without smoothing, so for a qualitative study of the overall velocity fields, the low-resolution ($3'' \times 3''$) smoothed map is used. However, the high-resolution map allows us to study the kinematics of the faintest regions.

\subsubsection{Rotation curve}

The rotation curve of the galaxy can be derived from the velocity map, which contains the projected velocity along the line-of-sight (v$_{\rm los}$) at each position of the galaxy. Only LOS velocities are observable so it is necessary to correct for projection effects. 

The standard method for deprojecting galaxies from kinematic data is to assume that the galaxy is made up of a number of annuli, the well-known tilted-ring model \citep{Begeman1989}, where each ring can be defined with the parameters: inclination ($i$), PA, centre ($x_{0}$ and $y_{0}$) and systemic velocity ($v_{sys}$). 

\begin{figure}
\begin{center}
\includegraphics[width=84mm]{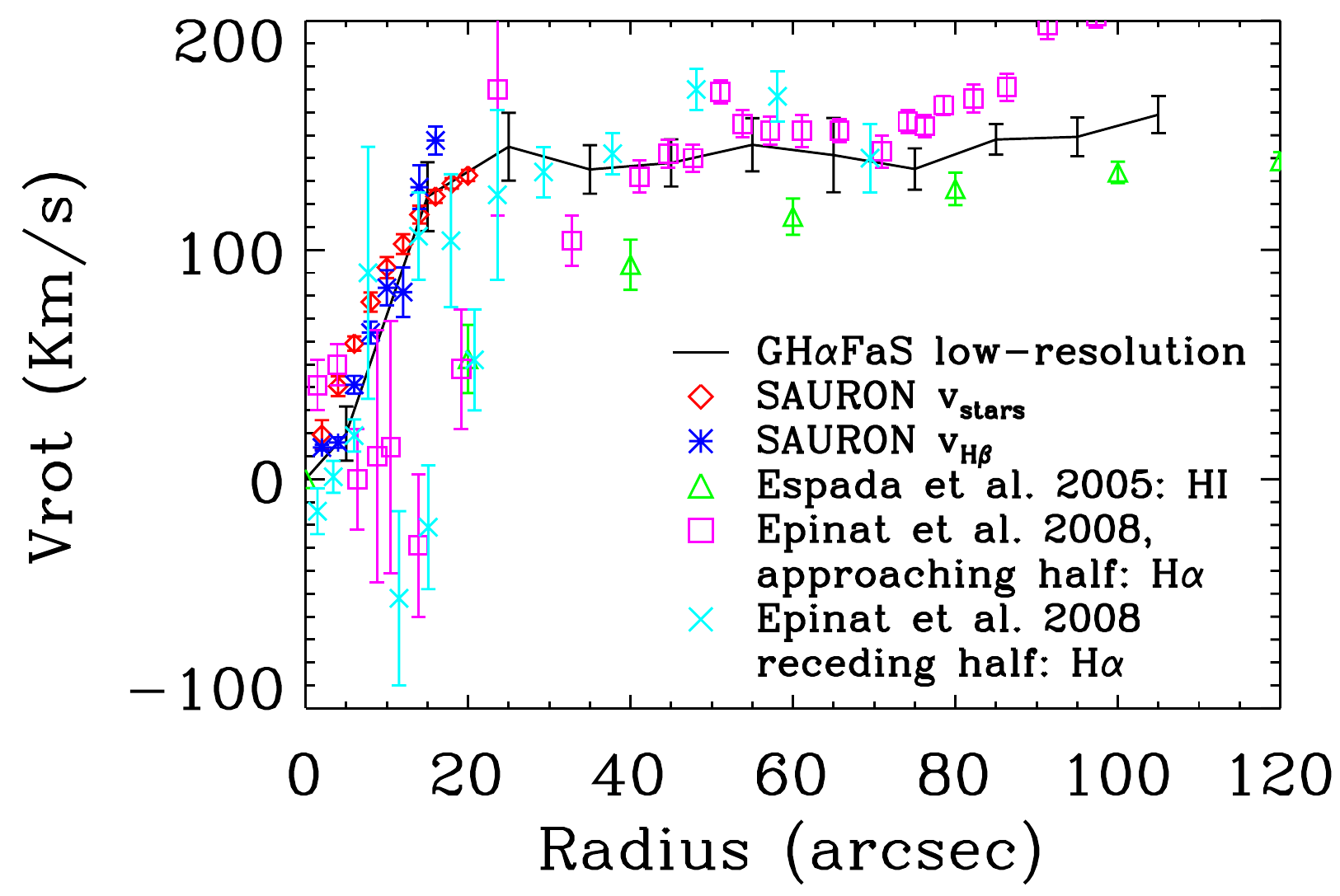}
\caption{Rotation curves for NGC 864. The black line corresponds to the rotation curve derived from the GH$\alpha$FaS data, using the low-resolution velocity map. The dark blue plus signs show the rotation curve derived from the SAURON H$\beta $ velocity map, whereas the red diamods show the rotation curve derived from the SAURON stellar velocity map. The green triangles correspond to the rotation curve from HI observations taken from \citet{Espada2005}. The pink squares and light blue crosses represent the rotation curves for the approaching and receding halves of NGC 864 derived in \citet{Epinat2008}. Note that the FOV of SAURON covers only the inner part of the galaxy.}
\label{ghafasRC}
\end{center}
\end{figure}

Rotation curves have been obtained with ROTCUR procedure in GIPSY, following the \citet{Begeman1989} method. Most initial entry values were taken from the literature, as in Table \ref{props}, but the initial rotation velocity values have been taken from the rotation curves in \citet{Espada2005}. Setting the expansion velocity to zero always, the following steps were carried out to obtain some galaxy parameters (galaxy centre, PA, \textit{i}) and the final rotation curve. (1) All parameters were left free. The resulting rotation curve was not correct and some parameters had to be fixed. (2) To obtain the position of the galaxy centre, the PA, inclination, systemic and rotation velocities were fixed, and the centre values were free. (3) The centre position, \textit{PA}, inclination, and rotation velocities were fixed in order to fit the systemic velocity. (4) The \textit{PA} and inclination were then left free with all other parameters fixed. (5) Finally, the rotation curve was obtained by fixing all parameters except for the rotation velocities. The final values for \textit{PA} and \textit{i} are presented in Table 1, whereas the resulting rotation curves can be found in Figure \ref{ghafasRC}. Note that the resulting rotation curves take into account the deprojected velocity using the computation of the different parameters. The same procedure has been followed to obtain the rotation curves derived from the SAURON stellar and gaseous velocity maps (Figure \ref{SAURONMAPS}), and the results are also presented in Figure \ref{ghafasRC}.

Prior to discussing the physical implications of the features in the observed rotation curve, we briefly discuss the rotation curves of NGC 864 found in the literature (from HI observations, \citealt{Espada2005}, reproduced in Figure \ref{ghafasRC} until 120 arcsec; and from H$ \alpha $ observations, \citealt{Epinat2008}) as compared to the rotation curves obtained using the GH$ \alpha $FaS FP data and the SAURON data. It is obvious that the information provided from each rotation curve is different, as each wavelength offers different information at different radial angles. On the one hand, the rotation curve derived from HI observations suffers from beam smearing due to the low spatial resolution, although it reaches the outer parts of the galaxy disc. On the other hand, rotation curves obtained from H$\alpha$ observations are based on detailed but also more patchy emission, which can make the analysis more difficult.

The complete HI rotation curve (Figure 7 in \citealt{Espada2005}) shows the behaviour that is typical of a galactic disc: solid-body rotation near the centre, and almost constant velocity in the outer parts. At approximately 100 arcsec from the centre ($\sim$ 10 kpc), the curve reaches a maximum of 130 km s$ ^{-1} $, and continues more or less at that velocity until the observed disc ends ($\sim$ 40.5 kpc).

The rotation curves derived from SAURON data show the first 20 arcseconds from the centre, which presents essentially solid-body rotation. However, the rotation curves derived from H$ \alpha $ observations extend as far as the optical disc, some 100 arcseconds from the centre ($\sim$ 10 kpc). The central parts show a quick rise in velocity, which is followed by an essentially flat rotation curve for R $ \ga 25''$. Although the receding and approaching halves have been analysed separately in the H$\alpha$ rotation curves found in the literature (pink squares and blue crosses in Figure \ref{ghafasRC}, taken from Figure E1 in Appendix 5 from \citealt{Epinat2008}), they show some common features when compared to the rotation curve from GH$ \alpha $FaS data: a dip of the curve in the central part of the galaxy, then a steep rise up to 30 arcsec from the centre (i.e., up to 3.04 kpc from the centre), and a less steep slope in the last part of the curve. Some of these features found in rotation curves from H$ \alpha $ observations can be a consequence of the reduction process (e.g., wrongly fixed parameters when deriving the rotation curve, such as the systemic velocity), or can be caused by a physical feature (e.g., the presence of a bar) or even due to the observations (patchiness in the line emission).

\subsubsection{Velocity models and residual maps}

The presence of non-circular motions in spiral galaxies can have a considerable impact on the shape of the rotation curve, and thus on the amount of dark matter inferred. On the other hand, the non-circular motions carry important information about the components that cause them. For instance, the strongly non-axisymmetric gravitational potential of the bar can cause local deviations on the velocity field (the effects of streaming motions on the gas flow in galaxies have been considered by e.g. \citealt{Duval1983} under the beam scheme; by \citealt{Lindblad1996} using the FS2 code presented in \citealt{vanAlbada1981,vanAlbada1982} and \citealt{vanAlbada1985}; or by \citealt{Salo1999} using the sticky-particle method), while strong spiral structure can provoke shocks and streaming motions in the gas (e.g., \citealt{Bosma1978};  \citealt{Visser1980}; \citealt{Marcelin1985}; \citealt{Rots1990}; \citealt{Knapen1993} and \citealt{Knapen2000}).

\begin{figure}
\begin{center}
 %\includegraphics[width=84mm]{modelheat.pdf}
%ONLINE VERSION:
 \includegraphics[width=84mm]{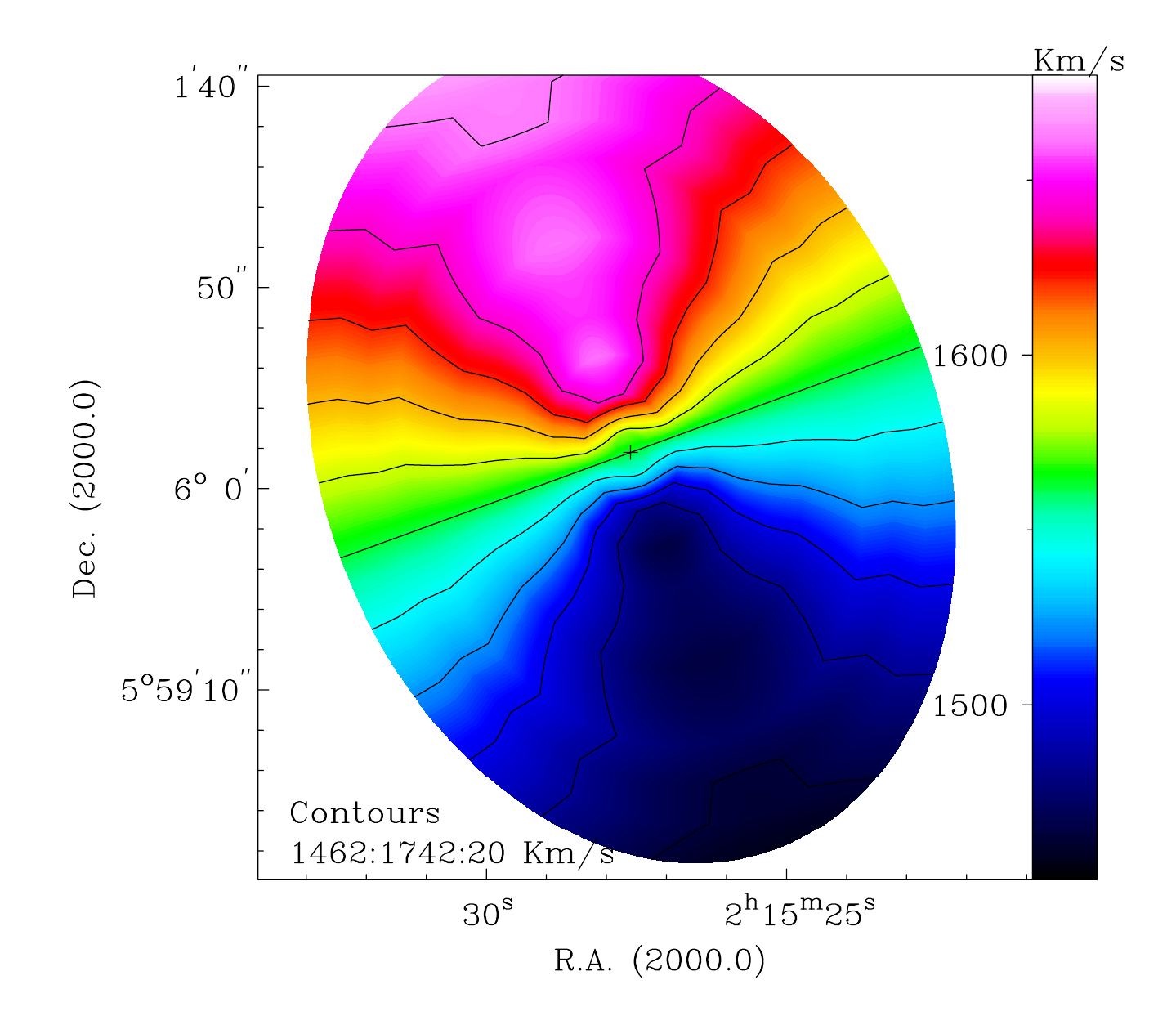}
\caption{Model map derived from the rotation velocities previously obtained in the rotation curve.}
\label{model}
\end{center}
\end{figure}

\begin{figure*}
\begin{center}
% \includegraphics[width=17cm]{restaheat.pdf}
%ONLINE VERSION:
 \includegraphics[width=17cm]{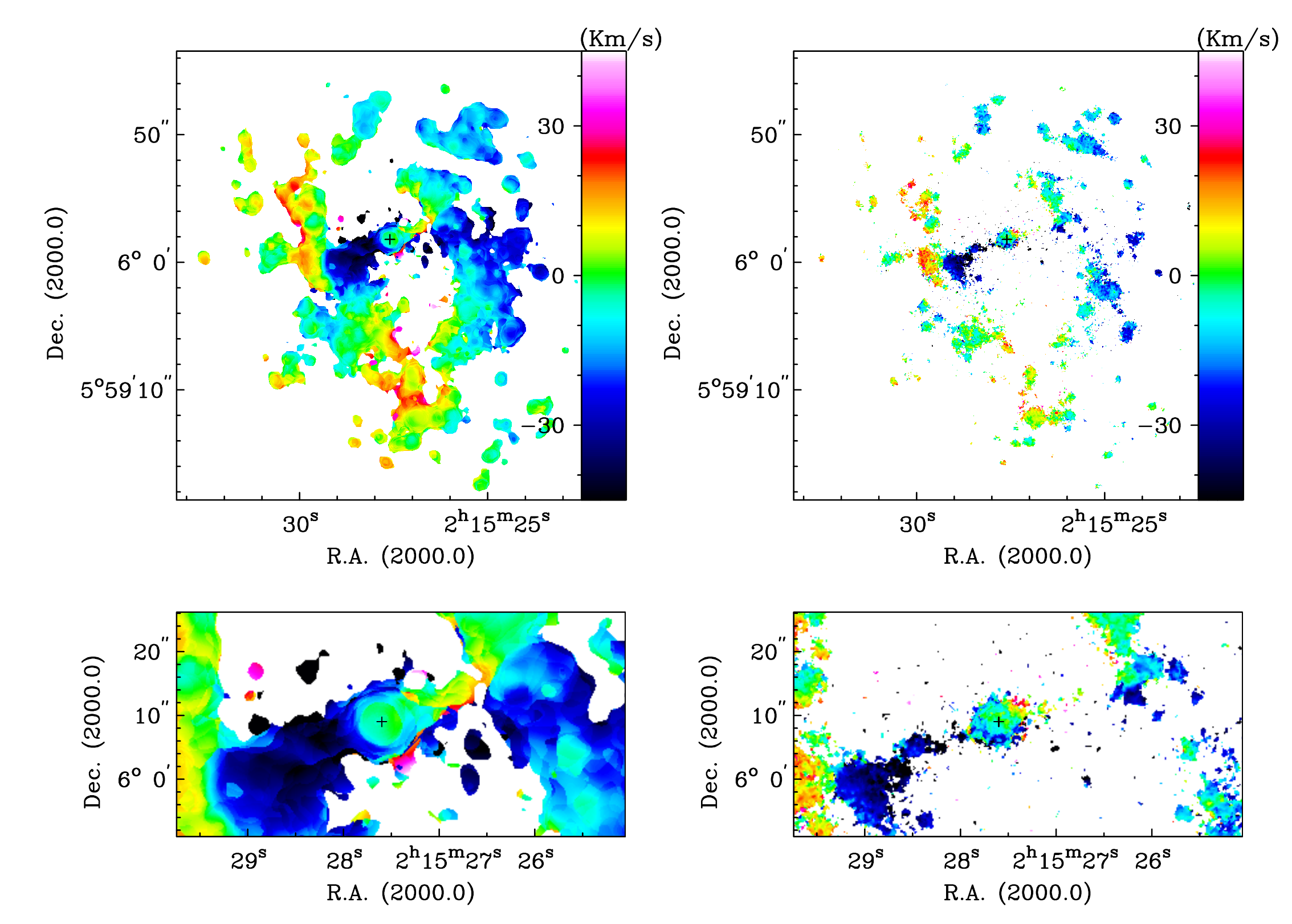}
\caption{Residual velocity maps obtained from the low-resolution map (\textit{left}) and from the high-resolution map (\textit{right}). \textit{Bottom}: the bar zone for both maps is highlighted.}
\label{resta}
\end{center}
\end{figure*}

To study the non-circular motions in NGC 864, we use the rotation curve derived above from the GH$ \alpha $FaS data to construct a model, assuming kinematic symmetry, and using the GIPSY task \textit{VELFI}. This procedure creates a model velocity field plotting the information from the provided rotation curve into the same dimensions as the original map (Figure \ref{model}). A residual velocity map (Figure \ref{resta}) is created by subtracting the model from the original velocity map. This residual map can be interpreted as a view of the non-circular motions.

These figures show that the region with the highest non-circular motions, of around 50 km/s in absolute value (about one third of the rotational velocity), is delineated by the bar. As explained before, it is easier to measure and highlight these non-circular motions in the smoothed map.

\subsubsection{Position-velocity diagrams}

One can study the non-circular motions by examining the residual velocity field, as we saw in the previous section. However, as each point on the rotation curve is calculated by azimuthally averaging the complete velocity information (including non-circular motions) within a certain range, the rotation curve and the resulting model velocity field will be influenced to some degree by non-circular motions. Consequently, we used position-velocity diagrams (PV) diagrams, because they reproduce all the observed information, even though they are not always easy to interpret.

\begin{figure}
\begin{center}
 \includegraphics[width=84mm]{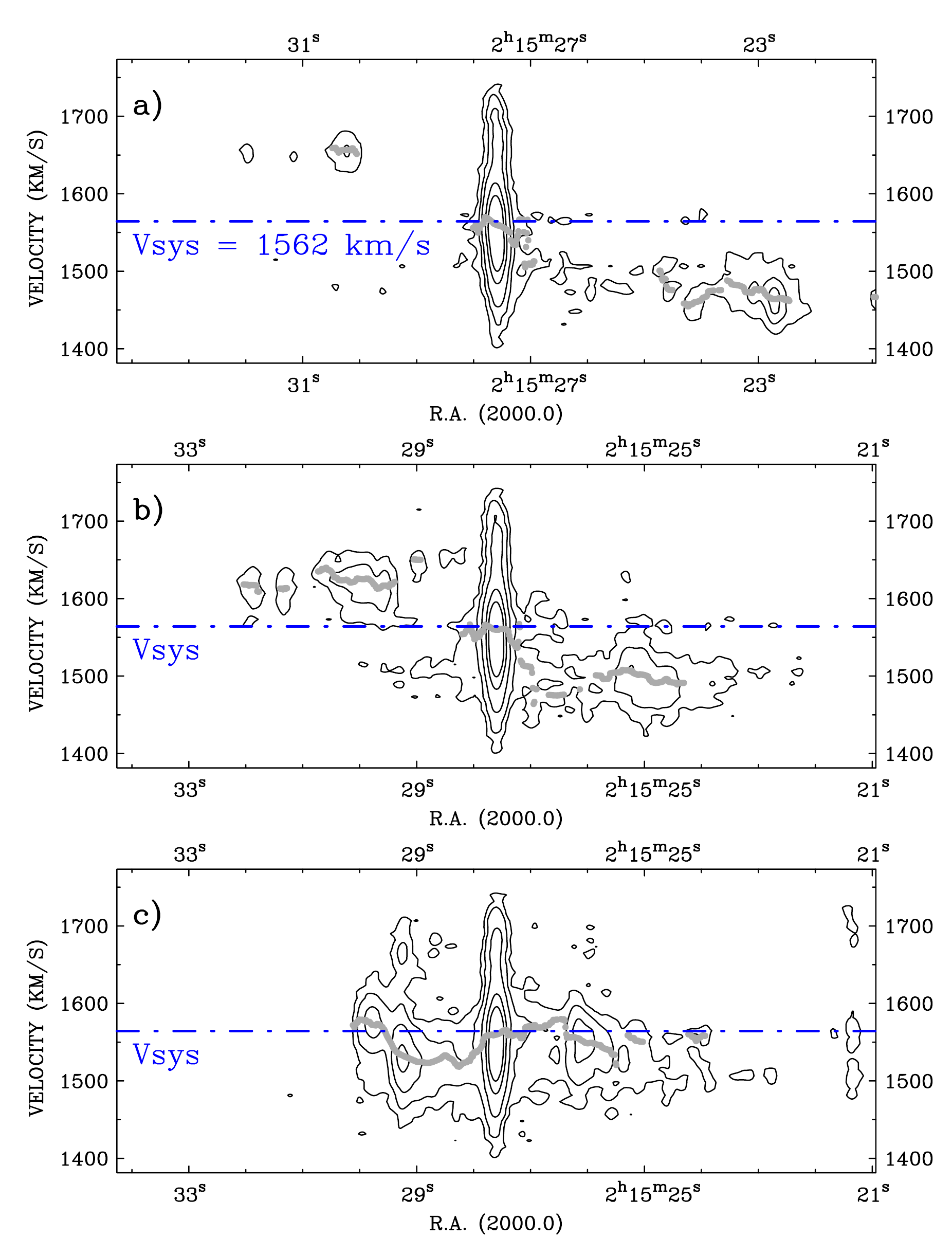}
\caption{PV diagrams for NGC 864 from the low-resolution GH$\alpha$FaS velocity data. \textit{a)} PV along the kinematic major axis (PA=20$\degr$, counter-clockwise from N); \textit{b)} PV along an intermediate angle (70$\degr$) and \textit{c)} PV along the kinematic minor axis (PA=110$\degr$). Contour levels are at 0.5, 1.5, 4.5, 13.5 and 25$\sigma$. Overlaid (grey dots) is the velocity profile for the whole disc at the corresponding angle and the systemic velocity (dash-dot-dash line) corresponding to a velocity of 1562 km/s.}
\label{PV}
\end{center}
\end{figure}

We used the \textit{SLICE} task in GIPSY to extract a plane from a data set at the angles of the kinematic major axis (20$\degr$), the kinematic minor axis (110$\degr$) and along an intermediate angle of 70$\degr$ (wedges of $ \pm $5$ \degr $ were used in all cases). The data used are the low-resolution GH$\alpha$FaS cube, and the resulting PV diagrams are presented in Figure \ref{PV}.

A PV diagram of a galaxy along the kinematic minor axis, in this case 110$\degr$, is completely flat in the absence of non-circular motions, whereas in the same ideal case a PV diagram along the kinematic major axis shows the rotation. At intermediate angles, such as the 70$\degr$ slice we show for NGC 864, an intermediate behaviour is expected. In the case we present here, the PV diagrams show clearly the presence of non-circular motions, and indicate their location and approximate amplitude. We will describe in the next section that they correspond to motions induced by the bar and spiral arms. The centre shows a very broad extent of velocities in H$ \alpha $ emission, suggesting a strong velocity gradient here. Additional contributions from strong star-forming activity in the nucleus, and/or incorrect continuum subtraction there, cannot be completely excluded. We will ignore the centre in our further analysis. The minor axis PV diagram shows a feature at 2$ ^{h} $15$ ^{m} $29$ ^{s} $ and $\sim$1660 km/s. It arises from the [NII]-6548 line leakage that didn't affect the computation of the moment maps, but that shows up in a PV diagram.

\subsection{Star formation rate}

 The star formation rate (SFR) was determined following \citet{Kennicutt2009} with the expression:
 
 \begin{equation}
 {\rm SFR} (M_{\odot}  {\rm yr^{-1}}) = 5.5 \times 10^{-42} L({\rm H\alpha}) 
 \end{equation}
 
 where $L({\rm H\alpha})$ is the luminosity, calculated as
 
 \begin{equation}
 L({\rm H\alpha}) [{\rm erg/s}]=4\pi D^{2} (3.086 \times 10^{24})^{2}F_{{\rm H\alpha}}^{*},
\end{equation}  

with \textit{D} being the distance to the galaxy in Mpc (Table 1) and $F_{{\rm H\alpha}}^{*}$ the flux corrected for Galactic and internal absorption. In this case, the Galactic absorption was taken from NED, with a value of $A(R)=0.130$ mag (calculated based on the dust maps from \citealt{Schlegel1998}). To correct for internal absorption, there are several proposals in the literature (see \citealt{Sanchez-Gallego2012} for an overview). We have adopted the constant value of $ A(H\alpha) = 1.1 $ mag (\citealt{Kennicutt1983}).

We have measured the SFR in the brightest regions [marked with letters in Fig. \ref{ghafasmaps}(a)], and also for the whole galaxy. The final values are presented in Table \ref{SFR}. Not all the H$ \alpha $ emission in the centre is necessarily due to massive star formation, with contributions possible from moderate non-stellar activiy (even though NGC 864 appears not to have an AGN activity, see \citealt{Ho1997}) or from shocks. This value should thus be considered as an upper limit, of around one tenth of the total SFR in the galaxy. The SFR for the bar region is low (0.03 M$ _{\odot} $ yr$ ^{-1} $), which can be explained by a dominant old population there. There is some massive star formation from the regions at the end of the bar, but almost all massive star formation in this galaxy occurs in the two main spiral arms.

\begin{table}

\caption{$L ({\rm H\alpha}) $ and SFR measured from the ACAM H$\alpha$ image for the whole galaxy, the eastern and western arms, for the bar region, and for the brightest regions named with A, B, C and D identified in panel \textit{a)} of Figure \ref{ghafasmaps} (the centre labelled with A; B in the south, at the end of the western arm, and C and D at the eastern and western ends of the bar, respectively).}
 \label{SFR}
\centering
\begin{tabular}{|c|c|c|}
\hline

Region  &  $ L ({\rm H\alpha}) $ (erg s$ ^{-1} $)  & SFR ($M _{\odot} $ yr$ ^{-1} $)  \\
\hline 
  Whole galaxy & $(4.0 \pm 1.6 )\cdot10 ^{41} $ &  2.19 $ \pm $ 0.88\\  
  Eastern arm & $(1.5 \pm 0.6 )\cdot10 ^{41} $ & 0.85 $ \pm $ 0.34\\
  Western arm & $(1.9 \pm 0.8 )\cdot10 ^{41} $ & 1.04 $ \pm $ 0.41\\ 
   Bar & $(5.3 \pm 2.1)\cdot10 ^{39} $ &  0.03 $ \pm $ 0.01\\
 A (centre) & $(3.7 \pm 1.5 )\cdot10 ^{40} $ & 0.20 $ \pm $ 0.08\\
  B (End of western arm) & $(2.0 \pm\ 0.8 )\cdot10 ^{40} $ & 0.11 $ \pm $ 0.04\\
  C (start of eastern arm) & $(1.0 \pm\ 0.4 )\cdot10 ^{40} $ & 0.06 $ \pm $ 0.02\\
  D (start of western arm) & $(3.8 \pm\ 1.5 )\cdot10 ^{39} $ & 0.02 $ \pm $ 0.01\\
    \hline 
\end{tabular}
\end{table} 

There are several factors to take into account to estimate the uncertainties in the measurements of the SFR. First of all, the adopted value for the galaxy distance taken from NED refers to \citet{Tully2009}, which used the Tully-Fisher relationship to compute the galaxy distance. Therefore, a typical 20\% uncertainty is estimated for the distance value. The basic reduction processes, i.e. flat-fielding correction, can contribute with around 2\% of uncertainty. Secondly, the zero-point calibration carries an uncertainty of 3\%, in agreement with the value of 2\% typical for photometric nights. The uncertainty in the flux measurement is around 1\%. However, apart from the uncertainty due to the distance, the highest uncertainty comes from the continuum-subtraction, estimated to be around 11\%. The scaling factor is somehow a subjective value, and to have an idea of the uncertainty, we have measured the scaling factors from two images: one slightly oversubtracted and another slightly undersubtracted, leading to an uncertainty of the 11\%. 

Altogether, taking into account the uncertainties related to the quality of the image (without considering the distance) we estimate an uncertainty of 17\%. The estimated uncertainty is comparable to previous studies using the same method: \citet{James2004} who found a value of around 14\%, \citet{Kennicutt2008} who found a value of 16\%, and \citet{Sanchez-Gallego2012} who found a value of 18 \%.

%%%%%%%%%%%%%%%%%%%%%%%%%%%%%%%%%%%%%%%%%%%%%%%%%%

\section{Discussion: influence of the bar on the kinematics}

\begin{figure}
\begin{center}
 %\includegraphics[width=84mm]{minoraxisheat.pdf}
 %ONLINE VERSION:
 \includegraphics[width=84mm]{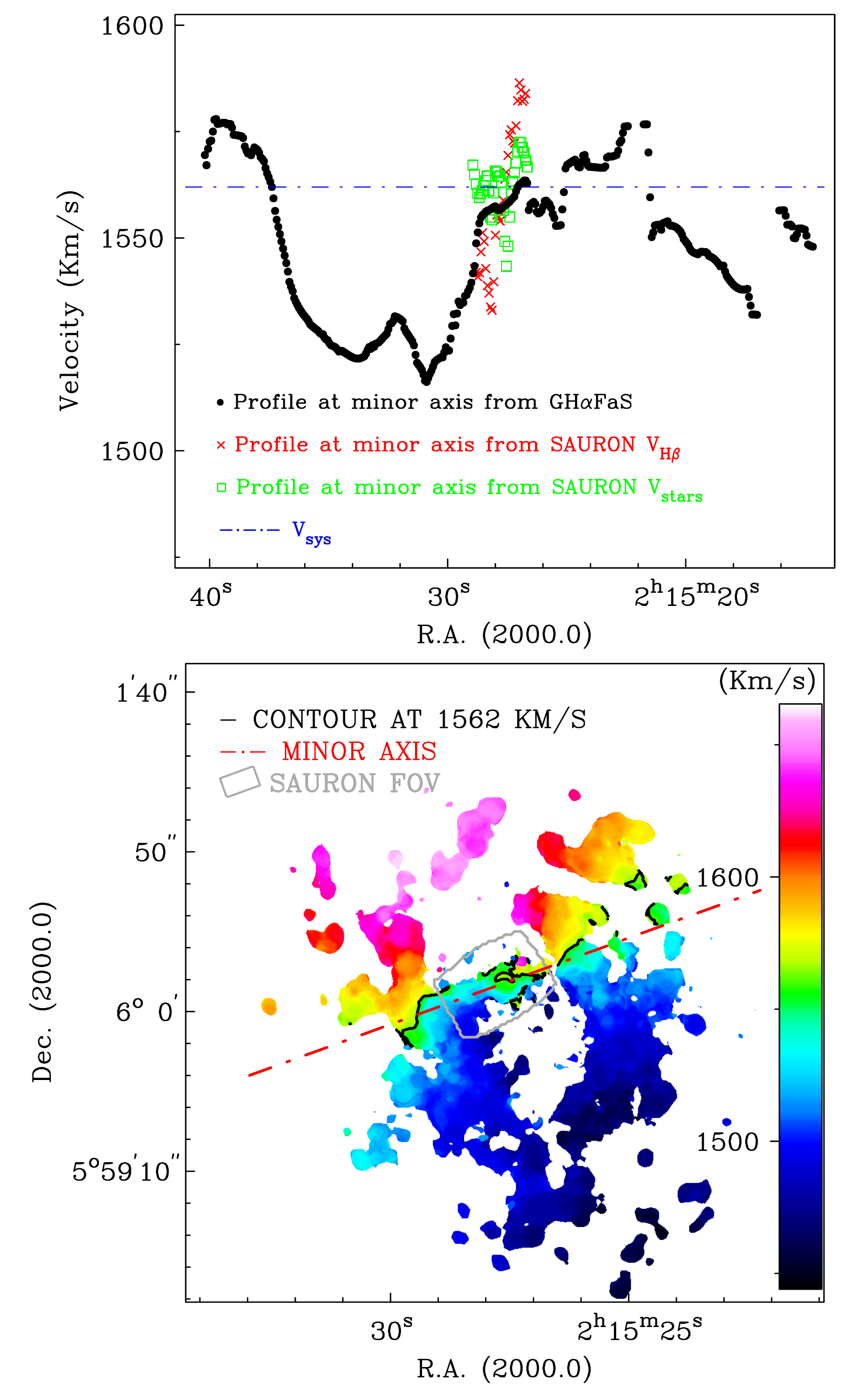}
\caption{\textit{Top}: the profile along the kinematic minor axis of NGC 864 in black dots; overlaid in a blue dash-dot-dash line the systemic velocity of 1562 km/s; overlaid in red crosses and green squares the 1562 km/s profiles at kinematic minor axis from the SAURON H$\beta$ velocity map and stellar velocity map. \textit{Bottom}: velocity map obtained with GIPSY. The red dash-dot-dashed line shows the location of the kinematic minor axis, whereas the black contours trace a velocity of 1562 km/s, corresponding to the systemic velocity. In grey is shown the location of the SAURON image to compare the different scales.}
\label{minoraxis}
\end{center}
\end{figure}

 The analysis of non-circular motions is of great interest to the understanding of galaxy structure and kinematics. The rotation curves (Figure \ref{ghafasRC}), PV diagrams (Figure \ref{PV}), and the residual maps (Figure \ref{resta}) provide crucial information about the kinematical processes that take place in the galaxy. A prominent bar, such as the one in NGC 864, is often the most important factor in this context, and indeed the presence of the non-axisymmetric gravitational potential of the bar can be recognised in the deviations from circular motion across the galaxy. In the residual maps (Figure \ref{resta}), the most prominent non-circular motions are found along the bar and  at the beginning of the eastern arm. This confirms that the bar has a significant influence on the kinematics of the galaxy, causing the velocities to deviate from the circular, rotational, motion. The kinematics along the kinematic minor axis may further quantify the influence of the potential of the bar, due to the fact that the PA of the bar is very similar to the PA of the kinematic minor axis. Figure \ref{minoraxis} thus presents the GH$\alpha$FaS velocity profile along the kinematic minor axis (top panel), and also the shape of the isovelocity contour at the systemic velocity (lower panel).

The profile shows that there are significant changes in the velocity along the kinematic minor axis, of up to 50 km/s. In the profile, it is noticeable that most of the velocities along the kinematic minor axis are below the systemic velocity. Various peaks and troughs are found along the profile. To study this further, avoiding the problem of the patchiness of the H$\alpha$ emission, we repeat this exercise with the SAURON H$\beta$ and stellar velocity maps. Here, the peaks and troughs can easily be distinguished, and the deviations from the systemic velocity also range up to 50 km/s.

\begin{figure}
\begin{center}
% \includegraphics[width=84mm]{SAURON_contoursheat.pdf}
 %ONLINE VERSION:
 \includegraphics[width=84mm]{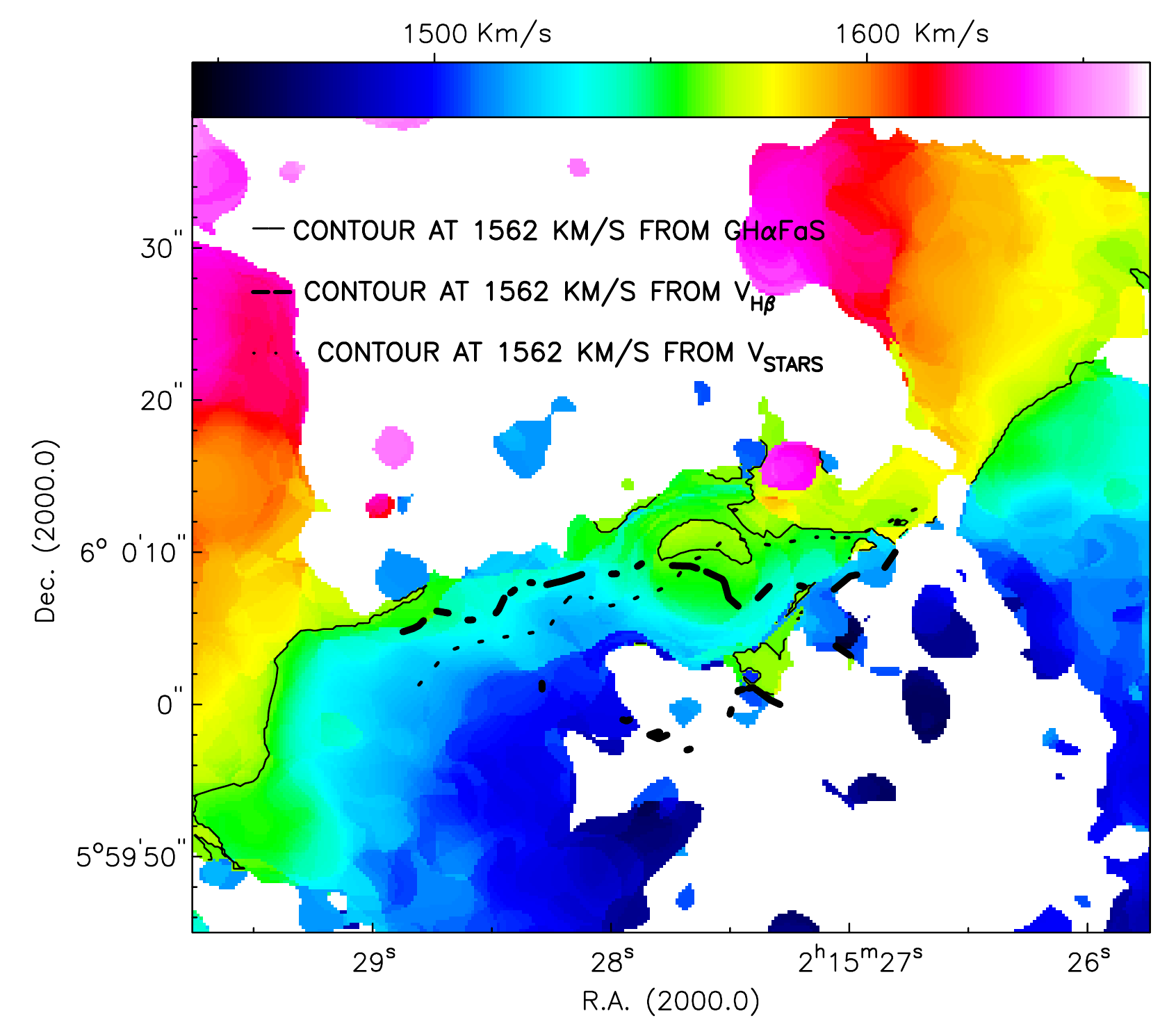}
\caption{Bar region of the low-resolution velocity map of NGC 864 from the GH$\alpha$FaS data. Overlaid are contours from SAURON data: the dashed thick line corresponds to a velocity of 1562 km s$^{-1}$ in the H$\beta$ velocity maps, whereas the dotted line corresponds to a velocity of 1562 km s$ ^{-1} $ in the stellar velocity map. The overlaid black thin line corresponds to 1562 km/s in GH$\alpha$FaS data.}
\label{SAURON_contours}
\end{center}
\end{figure}

One should note the differences between these SAURON maps and the one from GH$\alpha$FaS. Firstly, the FOV of SAURON (33$"  \times  41"$) is much smaller than the one of GH$\alpha$FaS ($202"  \times  202"$), so the profiles along the kinematic minor axis in this Figure are only the central part of the profile obtained with GH$\alpha$FaS data. Secondly, the patchiness implies a lack of points in some parts of the GH$\alpha$FaS profile. In Figure \ref{SAURON_contours}, we present the bar region of NGC 864 as seen in the low-resolution velocity map from GH$\alpha$FaS data, and overlaid, the intensity contours from SAURON H$\beta$ observations and the contours of the derived systemic velocity from SAURON H$\beta$ and stellar velocity maps.

In the profile from the GH$\alpha$FaS velocity map, there are lower velocities in the eastern part of the bar, and higher velocities in the western part. PV diagrams (Figure \ref{PV}) confirm this. The PV diagram along the kinematic major axis shows the rotation, whereas the PV diagram along the kinematic minor axis confirms the deviations along the bar with the peaks mentioned above. The correspondence between peaks can indicate bar streaming motions, although these results are not nearly as conclusive as in, for example, the much clearer case of M100, as described in, e.g., \citet{Knapen2000}.

Taking everything into account, the profiles and PV diagrams along the kinematic minor axis show that there are deviations from the circular rotation velocity, confirming the presence of non-circular motions along the bar. These are probably caused by the non-axisymmetrical potential created by the bar, as widely discussed in the literature (see references in Section 4.2.2).

%%%%%%%%%%%%%%%%%%%%%%%%%%%%%%%%%%%%%%%%%%%%%%%%%%

\section{Conclusions}

We have presented a kinematic study of the spiral galaxy NGC 864. Data sets obtained with three different WHT instruments have been studied, resulting in rotation curves and intensity, velocity, residual and gradient maps. From this, we reach the following conclusions:

\begin{enumerate}
\item A technique for flux calibration of GH$ \alpha $FaS data has been developed. This will be used for other S$ ^{4} $G galaxies observed with GH$ \alpha $FaS.
\item Our bulge/disc/bar 2D decomposition of the S$^4$G 3.6$\mu m$ image of NGC 864 using {\sc budda} yields a flat-profile bar, which is typical of strong bars (\citealt{Athanassoula2002}) and of two-armed grand design spirals (\citealt{Elmegreen1985}). Flat-profile bars are linked to a higher degree of exchange of angular momentum with the halo than exponential bars (\citealt{Athanassoula2003}). Such bars can result from sharply rising rotation curves, as observed in NGC 864, and are most likely connected to the observed spatial offset between the ends of the bars and the onset of the spiral arms.
\item The rotation curves obtained from the GH$ \alpha $FaS velocity maps are similar to those obtained by \citet{Epinat2008}, both using H$ \alpha $ observations. However, the curves do not cover all the disc and the intrinsic patchiness linked to the line emission prejudices the analysis and the derived results. Due to the high angular and spectral resolution, our new rotation curve shows considerable detail in the central 2 kpc radius.
\item We have found non-circular motions along the bar in the residual maps, in the PV diagram along the kinematic minor axis, and in the velocity profile along the kinematic minor axis. These velocity patterns are typical of a barred galaxy behaviour, with streaming motions along the bar.

\end{enumerate}

This paper is the first in a series presenting the results of the programme to observe a sample of some 40 spiral galaxies of all morphological types from the S$ ^{4} $G sample with the GH$ \alpha $FaS FP instrument.

\section*{Acknowledgments}
We acknowledge financial support to the DAGAL network from the People Programme (Marie Curie Actions) of the European Union's Seventh Framework Programme FP7/2007-2013/ under REA grant agreement number PITN-GA-2011-289313. This work was co-funded under the Marie Curie Actions of the European Commission (FP7-COFUND). We also gratefully acknowledge support from NASA JPL/Spitzer grant RSA 1374189 provided for the S$ ^{4} $G project. E.A. and A.B. thank the CNES for support. K.S., J.-C.M.-M., T.K., and T.M. acknowledge support from the National Radio Astronomy Observatory, which is a facility of the National Science Foundation operated under cooperative agreement by Associated Universities, Inc. JFB acknowledges support from the Ram\'on y Cajal programme as well as grant AYA2010-21322-C03-02 by the Spanish Ministry of Economy and Competitiveness (MINECO). This research has been supported by the MINECO under the grant AYA2007-67625-CO2-O2. This work is based on observations made with the WHT operated on the island of La Palma by the Isaac Newton Group of Telescopes, in the Spanish Observatorio del Roque de Los Muchachos of the Instituto de Astrof\'isica de Canarias. The authors thank the entire S$ ^{4} $G team for their efforts in this project. This work is based on observations and archival data made with the \textit{Spitzer Space Telescope}, which is operated by the Jet Propulsion Laboratory, California Institute of Technology under a contract with NASA. We are grateful to the dedicated staff at the \textit{Spitzer} Science Center for their help and support in planning and execution of this Exploration Science programme. This research has made use of the NASA/IPAC Extragalactic Database (NED) which is operated by JPL, Caltech, under contract with NASA.
%%%%%%%%%%%%%%%%%%%%%%%%%%%%%%%%%%%%%%%%%%%%%%%%%%

\bibliographystyle{mn2e}
\bibliography{references}

\begin{thebibliography}{95}
\expandafter\ifx\csname natexlab\endcsname\relax\def\natexlab#1{#1}\fi

\bibitem[{{Amram} {et~al}\mbox{.}(1992){Amram}, {Le Coarer}, {Marcelin},
  {Balkowski}, {Sullivan}, \& {Cayatte}}]{Amram1992}
{Amram} P., {Le Coarer} E., {Marcelin} M., {Balkowski} C., {Sullivan}, III
  W.~T., {Cayatte} V., 1992, \aaps, 94, 175

\bibitem[{{Amram} {et~al}\mbox{.}(1994){Amram}, {Marcelin}, {Balkowski},
  {Cayatte}, {Sullivan}, \& {Le Coarer}}]{Amram1994}
{Amram} P., {Marcelin} M., {Balkowski} C., {Cayatte} V., {Sullivan}, III W.~T.,
  {Le Coarer} E., 1994, \aaps, 103, 5

\bibitem[{{Athanassoula}(2003)}]{Athanassoula2003}
{Athanassoula} E., 2003, \mnras, 341, 1179

\bibitem[{{Athanassoula} \& {Beaton}(2006)}]{Athanassoula2006}
{Athanassoula} E., {Beaton} R.~L., 2006, \mnras, 370, 1499

\bibitem[{{Athanassoula} {et~al}\mbox{.}(2009){Athanassoula}, {Gadotti},
  {Carrasco}, {Bosma}, {de Souza}, \& {Recillas}}]{Athanassoula2009a}
{Athanassoula} E., {Gadotti} D.~A., {Carrasco} L., {Bosma} A., {de Souza}
  R.~E., {Recillas} E., 2009, in Revista Mexicana de Astronomia y Astrofisica,
  vol. 27, Vol.~37, Revista Mexicana de Astronomia y Astrofisica Conference
  Series, pp. 79--82

\bibitem[{{Athanassoula} \& {Misiriotis}(2002)}]{Athanassoula2002}
{Athanassoula} E., {Misiriotis} A., 2002, \mnras, 330, 35

\bibitem[{{Bacon} {et~al}\mbox{.}(2001){Bacon}, {Copin}, {Monnet}, {Miller},
  {Allington-Smith}, {Bureau}, {Carollo}, {Davies}, {Emsellem}, {Kuntschner},
  {Peletier}, {Verolme}, \& {de Zeeuw}}]{Bacon2001}
{Bacon} R. {et~al.}, 2001, \mnras, 326, 23

\bibitem[{{Begeman}(1989)}]{Begeman1989}
{Begeman} K.~G., 1989, \aap, 223, 47

\bibitem[{{Benn}, {Dee} \& {Ag{\'o}cs}(2008){Benn}, {Dee}, \&
  {Ag{\'o}cs}}]{Benn2008}
{Benn} C., {Dee} K., {Ag{\'o}cs} T., 2008, in Presented at the Society of
  Photo-Optical Instrumentation Engineers (SPIE) Conference, Vol. 7014, Society
  of Photo-Optical Instrumentation Engineers (SPIE) Conference Series

\bibitem[{{Block}, {Elmegreen} \& {Wainscoat}(1996){Block}, {Elmegreen}, \&
  {Wainscoat}}]{Block1996}
{Block} D.~L., {Elmegreen} B.~G., {Wainscoat} R.~J., 1996, \nat, 381, 674

\bibitem[{{Bonnarel} {et~al}\mbox{.}(1988){Bonnarel}, {Boulesteix},
  {Georgelin}, {Lecoarer}, {Marcelin}, {Bacon}, \& {Monnet}}]{Bonnarel1988}
{Bonnarel} F., {Boulesteix} J., {Georgelin} Y.~P., {Lecoarer} E., {Marcelin}
  M., {Bacon} R., {Monnet} G., 1988, \aap, 189, 59

\bibitem[{{Bosma}(1978)}]{Bosma1978}
{Bosma} A., 1978, PhD thesis, PhD Thesis, Groningen Univ., (1978)

\bibitem[{{Bosma}(1981)}]{Bosma1981}
---, 1981, \aj, 86, 1791

\bibitem[{{Buta} {et~al}\mbox{.}(2005){Buta}, {Vasylyev}, {Salo}, \&
  {Laurikainen}}]{Buta2005}
{Buta} R., {Vasylyev} S., {Salo} H., {Laurikainen} E., 2005, \aj, 130, 506

\bibitem[{{Buta}, {Corwin} \& {Odewahn}(2007){Buta}, {Corwin}, \&
  {Odewahn}}]{Buta2007}
{Buta} R.~J., {Corwin} H.~G., {Odewahn} S.~C., 2007, {The de Vaucouleurs Altlas
  of Galaxies}, {Buta, R.~J., Corwin, H.~G., \& Odewahn, S.~C.}, ed. Cambridge
  University Press

\bibitem[{{Buta} \& {Zhang}(2009)}]{Buta2009}
{Buta} R.~J., {Zhang} X., 2009, \apjs, 182, 559

\bibitem[{{Carignan} {et~al}\mbox{.}(2008){Carignan}, {Hernandez}, {Beckman},
  \& {Fathi}}]{Carignan2008}
{Carignan} C., {Hernandez} O., {Beckman} J.~E., {Fathi} K., 2008, in
  Astronomical Society of the Pacific Conference Series, Vol. 390, Pathways
  Through an Eclectic Universe, {J.~H.~Knapen, T.~J.~Mahoney, \& A.~Vazdekis},
  ed., pp. 168--+

\bibitem[{{Cecil}, {Wilson} \& {Tully}(1992){Cecil}, {Wilson}, \&
  {Tully}}]{Cecil1992}
{Cecil} G., {Wilson} A.~S., {Tully} R.~B., 1992, \apj, 390, 365

\bibitem[{{Christlein} \& {Zaritsky}(2008)}]{Christlein2008}
{Christlein} D., {Zaritsky} D., 2008, \apj, 680, 1053

\bibitem[{{Combes} \& {Elmegreen}(1993)}]{Combes1993}
{Combes} F., {Elmegreen} B.~G., 1993, \aap, 271, 391

\bibitem[{{Comer{\'o}n} {et~al}\mbox{.}(2010){Comer{\'o}n}, {Knapen},
  {Beckman}, {Laurikainen}, {Salo}, {Mart{\'{\i}}nez-Valpuesta}, \&
  {Buta}}]{Comeron2010}
{Comer{\'o}n} S., {Knapen} J.~H., {Beckman} J.~E., {Laurikainen} E., {Salo} H.,
  {Mart{\'{\i}}nez-Valpuesta} I., {Buta} R.~J., 2010, \mnras, 402, 2462

\bibitem[{{Corradi} \& {Capaccioli}(1991)}]{Corradi1991}
{Corradi} R.~L.~M., {Capaccioli} M., 1991, \aaps, 90, 121

\bibitem[{{de Souza}, {Gadotti} \& {dos Anjos}(2004){de Souza}, {Gadotti}, \&
  {dos Anjos}}]{Souza2004}
{de Souza} R.~E., {Gadotti} D.~A., {dos Anjos} S., 2004, \apjs, 153, 411

\bibitem[{{de Vaucouleurs} {et~al}\mbox{.}(1991){de Vaucouleurs}, {de
  Vaucouleurs}, {Corwin}, {Buta}, {Paturel}, \& {Fouque}}]{RC3}
{de Vaucouleurs} G., {de Vaucouleurs} A., {Corwin}, Jr. H.~G., {Buta} R.~J.,
  {Paturel} G., {Fouque} P., 1991, {Third Reference Catalogue of Bright
  Galaxies}, {de Vaucouleurs, G., de Vaucouleurs, A., Corwin, H.~G., Jr., Buta,
  R.~J., Paturel, G., \& Fouque, P.}, ed.

\bibitem[{{de Vaucouleurs} \& {Pence}(1980)}]{deVaucouleurs1980}
{de Vaucouleurs} G., {Pence} W.~D., 1980, \apj, 242, 18

\bibitem[{{Deharveng} \& {Pellet}(1975)}]{Deharveng1975}
{Deharveng} J.~M., {Pellet} A., 1975, \aap, 38, 15

\bibitem[{{Dubout} {et~al}\mbox{.}(1976){Dubout}, {Laval}, {Maucherat},
  {Monnet}, {Petit}, \& {Simien}}]{Dubout1976}
{Dubout} R., {Laval} A., {Maucherat} A., {Monnet} G., {Petit} M., {Simien} F.,
  1976, \aplett, 17, 141

\bibitem[{{Duval} \& {Athanassoula}(1983)}]{Duval1983}
{Duval} M.~F., {Athanassoula} E., 1983, \aap, 121, 297

\bibitem[{{Elmegreen} \& {Elmegreen}(1985)}]{Elmegreen1985}
{Elmegreen} B.~G., {Elmegreen} D.~M., 1985, \apj, 288, 438

\bibitem[{{Elmegreen} \& {Elmegreen}(1987)}]{Elmegreen1987}
{Elmegreen} D.~M., {Elmegreen} B.~G., 1987, \apj, 314, 3

\bibitem[{{Elmegreen} {et~al}\mbox{.}(2011){Elmegreen}, {Elmegreen}, {Yau},
  {Athanassoula}, {Bosma}, {Buta}, {Helou}, {Ho}, {Gadotti}, {Knapen},
  {Laurikainen}, {Madore}, {Masters}, {Meidt}, {Men{\'e}ndez-Delmestre},
  {Regan}, {Salo}, {Sheth}, {Zaritsky}, {Aravena}, {Skibba}, {Hinz}, {Laine},
  {Gil de Paz}, {Mu{\~n}oz-Mateos}, {Seibert}, {Mizusawa}, {Kim}, \& {Erroz
  Ferrer}}]{Elmegreen2011}
{Elmegreen} D.~M. {et~al.}, 2011, \apj, 737, 32

\bibitem[{{Epinat}, {Amram} \& {Marcelin}(2008){Epinat}, {Amram}, \&
  {Marcelin}}]{Epinat2008}
{Epinat} B., {Amram} P., {Marcelin} M., 2008, \mnras, 390, 466

\bibitem[{{Eskridge} {et~al}\mbox{.}(2000){Eskridge}, {Frogel}, {Pogge},
  {Quillen}, {Davies}, {DePoy}, {Houdashelt}, {Kuchinski}, {Ram{\'{\i}}rez},
  {Sellgren}, {Terndrup}, \& {Tiede}}]{Eskridge2000}
{Eskridge} P.~B. {et~al.}, 2000, \aj, 119, 536

\bibitem[{{Espada} {et~al}\mbox{.}(2005){Espada}, {Bosma}, {Verdes-Montenegro},
  {Athanassoula}, {Leon}, {Sulentic}, \& {Yun}}]{Espada2005}
{Espada} D., {Bosma} A., {Verdes-Montenegro} L., {Athanassoula} E., {Leon} S.,
  {Sulentic} J., {Yun} M.~S., 2005, \aap, 442, 455

\bibitem[{{Fazio} {et~al}\mbox{.}(2004){Fazio}, {Hora}, {Allen}, {Ashby},
  {Barmby}, {Deutsch}, {Huang}, {Kleiner}, {Marengo}, {Megeath}, {Melnick},
  {Pahre}, {Patten}, {Polizotti}, {Smith}, {Taylor}, {Wang}, {Willner},
  {Hoffmann}, {Pipher}, {Forrest}, {McMurty}, {McCreight}, {McKelvey},
  {McMurray}, {Koch}, {Moseley}, {Arendt}, {Mentzell}, {Marx}, {Losch},
  {Mayman}, {Eichhorn}, {Krebs}, {Jhabvala}, {Gezari}, {Fixsen}, {Flores},
  {Shakoorzadeh}, {Jungo}, {Hakun}, {Workman}, {Karpati}, {Kichak}, {Whitley},
  {Mann}, {Tollestrup}, {Eisenhardt}, {Stern}, {Gorjian}, {Bhattacharya},
  {Carey}, {Nelson}, {Glaccum}, {Lacy}, {Lowrance}, {Laine}, {Reach},
  {Stauffer}, {Surace}, {Wilson}, {Wright}, {Hoffman}, {Domingo}, \&
  {Cohen}}]{Fazio2004}
{Fazio} G.~G. {et~al.}, 2004, \apjs, 154, 10

\bibitem[{{Gadotti}(2008)}]{Gadotti2008}
{Gadotti} D.~A., 2008, \mnras, 384, 420

\bibitem[{{Gadotti} {et~al}\mbox{.}(2007){Gadotti}, {Athanassoula}, {Carrasco},
  {Bosma}, {de Souza}, \& {Recillas}}]{Gadotti2007}
{Gadotti} D.~A., {Athanassoula} E., {Carrasco} L., {Bosma} A., {de Souza}
  R.~E., {Recillas} E., 2007, \mnras, 381, 943

\bibitem[{{Ganda} {et~al}\mbox{.}(2006){Ganda}, {Falc{\'o}n-Barroso},
  {Peletier}, {Cappellari}, {Emsellem}, {McDermid}, {de Zeeuw}, \&
  {Carollo}}]{Ganda2006}
{Ganda} K., {Falc{\'o}n-Barroso} J., {Peletier} R.~F., {Cappellari} M.,
  {Emsellem} E., {McDermid} R.~M., {de Zeeuw} P.~T., {Carollo} C.~M., 2006,
  \mnras, 367, 46

\bibitem[{{Garrido} {et~al}\mbox{.}(2002){Garrido}, {Marcelin}, {Amram}, \&
  {Boulesteix}}]{Garrido2002}
{Garrido} O., {Marcelin} M., {Amram} P., {Boulesteix} J., 2002, \aap, 387, 821

\bibitem[{{Gottesman}(1982)}]{Gottesman1982}
{Gottesman} S.~T., 1982, \aj, 87, 751

\bibitem[{{Ho}, {Filippenko} \& {Sargent}(1997){Ho}, {Filippenko}, \&
  {Sargent}}]{Ho1997}
{Ho} L.~C., {Filippenko} A.~V., {Sargent} W.~L.~W., 1997, \apj, 487, 591

\bibitem[{{Hunt} \& {Malkan}(1999)}]{Hunt1999}
{Hunt} L.~K., {Malkan} M.~A., 1999, \apj, 516, 660

\bibitem[{{James} {et~al}\mbox{.}(2004){James}, {Shane}, {Beckman}, {Cardwell},
  {Collins}, {Etherton}, {de Jong}, {Fathi}, {Knapen}, {Peletier}, {Percival},
  {Pollacco}, {Seigar}, {Stedman}, \& {Steele}}]{James2004}
{James} P.~A. {et~al.}, 2004, \aap, 414, 23

\bibitem[{{James} {et~al}\mbox{.}(2005){James}, {Shane}, {Knapen}, {Etherton},
  \& {Percival}}]{James2005}
{James} P.~A., {Shane} N.~S., {Knapen} J.~H., {Etherton} J., {Percival} S.~M.,
  2005, \aap, 429, 851

\bibitem[{{Jim{\'e}nez-Vicente} {et~al}\mbox{.}(1999){Jim{\'e}nez-Vicente},
  {Battaner}, {Rozas}, {Casta{\~n}eda}, \& {Porcel}}]{Jimenez-Vicente1999}
{Jim{\'e}nez-Vicente} J., {Battaner} E., {Rozas} M., {Casta{\~n}eda} H.,
  {Porcel} C., 1999, \aap, 342, 417

\bibitem[{{Karachentseva}(1973)}]{Karachentseva1973}
{Karachentseva} V.~E., 1973, Astrofizicheskie Issledovaniia Izvestiya
  Spetsial'noj Astrofizicheskoj Observatorii, 8, 3

\bibitem[{{Kennicutt} {et~al}\mbox{.}(2009){Kennicutt}, {Hao}, {Calzetti},
  {Moustakas}, {Dale}, {Bendo}, {Engelbracht}, {Johnson}, \&
  {Lee}}]{Kennicutt2009}
{Kennicutt}, Jr. R.~C. {et~al.}, 2009, \apj, 703, 1672

\bibitem[{{Kennicutt} \& {Kent}(1983)}]{Kennicutt1983}
{Kennicutt}, Jr. R.~C., {Kent} S.~M., 1983, \aj, 88, 1094

\bibitem[{{Kennicutt} {et~al}\mbox{.}(2008){Kennicutt}, {Lee}, {Funes},
  {Sakai}, \& {Akiyama}}]{Kennicutt2008}
{Kennicutt}, Jr. R.~C., {Lee} J.~C., {Funes}, Jos{\'e}~G. S.~J., {Sakai} S.,
  {Akiyama} S., 2008, \apjs, 178, 247

\bibitem[{{Knapen} {et~al}\mbox{.}(1995){Knapen}, {Beckman}, {Heller},
  {Shlosman}, \& {de Jong}}]{Knapen1995}
{Knapen} J.~H., {Beckman} J.~E., {Heller} C.~H., {Shlosman} I., {de Jong}
  R.~S., 1995, \apj, 454, 623

\bibitem[{{Knapen} {et~al}\mbox{.}(1993){Knapen}, {Cepa}, {Beckman}, {Soledad
  del Rio}, \& {Pedlar}}]{Knapen1993}
{Knapen} J.~H., {Cepa} J., {Beckman} J.~E., {Soledad del Rio} M., {Pedlar} A.,
  1993, \apj, 416, 563

\bibitem[{{Knapen}, {Shlosman} \& {Peletier}(2000){Knapen}, {Shlosman}, \&
  {Peletier}}]{Knapen2000}
{Knapen} J.~H., {Shlosman} I., {Peletier} R.~F., 2000, \apj, 529, 93

\bibitem[{{Knapen} {et~al}\mbox{.}(2004){Knapen}, {Stedman}, {Bramich},
  {Folkes}, \& {Bradley}}]{Knapen2004}
{Knapen} J.~H., {Stedman} S., {Bramich} D.~M., {Folkes} S.~L., {Bradley} T.~R.,
  2004, \aap, 426, 1135

\bibitem[{{Kormendy} \& {Kennicutt}(2004)}]{Kormendy2004}
{Kormendy} J., {Kennicutt}, Jr. R.~C., 2004, \araa, 42, 603

\bibitem[{{Laine} {et~al}\mbox{.}(2002){Laine}, {Shlosman}, {Knapen}, \&
  {Peletier}}]{Laine2002}
{Laine} S., {Shlosman} I., {Knapen} J.~H., {Peletier} R.~F., 2002, \apj, 567,
  97

\bibitem[{{Laurikainen}, {Salo} \& {Buta}(2004){Laurikainen}, {Salo}, \&
  {Buta}}]{Laurikainen2004a}
{Laurikainen} E., {Salo} H., {Buta} R., 2004, \apj, 607, 103

\bibitem[{{Laurikainen} {et~al}\mbox{.}(2009){Laurikainen}, {Salo}, {Buta}, \&
  {Knapen}}]{Laurikainen2009}
{Laurikainen} E., {Salo} H., {Buta} R., {Knapen} J.~H., 2009, \apjl, 692, L34

\bibitem[{{Leon} \& {Verdes-Montenegro}(2003)}]{Leon2003}
{Leon} S., {Verdes-Montenegro} L., 2003, \aap, 411, 391

\bibitem[{{Lindblad}, {Lindblad} \& {Athanassoula}(1996){Lindblad}, {Lindblad},
  \& {Athanassoula}}]{Lindblad1996}
{Lindblad} P.~A.~B., {Lindblad} P.~O., {Athanassoula} E., 1996, \aap, 313, 65

\bibitem[{{Marcelin}, {Boulesteix} \& {Georgelin}(1985){Marcelin},
  {Boulesteix}, \& {Georgelin}}]{Marcelin1985}
{Marcelin} M., {Boulesteix} J., {Georgelin} Y.~P., 1985, \aap, 151, 144

\bibitem[{{Marinova} \& {Jogee}(2007)}]{Marinova2007}
{Marinova} I., {Jogee} S., 2007, \apj, 659, 1176

\bibitem[{{Martinez-Valpuesta}, {Knapen} \& {Buta}(2007){Martinez-Valpuesta},
  {Knapen}, \& {Buta}}]{Martinez-Valpuesta2007}
{Martinez-Valpuesta} I., {Knapen} J.~H., {Buta} R., 2007, \aj, 134, 1863

\bibitem[{{Martini} {et~al}\mbox{.}(2003){Martini}, {Regan}, {Mulchaey}, \&
  {Pogge}}]{Martini2003}
{Martini} P., {Regan} M.~W., {Mulchaey} J.~S., {Pogge} R.~W., 2003, \apjs, 146,
  353

\bibitem[{{Mathewson} \& {Ford}(1996)}]{Mathewson1996}
{Mathewson} D.~S., {Ford} V.~L., 1996, \apjs, 107, 97

\bibitem[{{Mathewson}, {Ford} \& {Buchhorn}(1992){Mathewson}, {Ford}, \&
  {Buchhorn}}]{Mathewson1992}
{Mathewson} D.~S., {Ford} V.~L., {Buchhorn} M., 1992, \apjs, 81, 413

\bibitem[{{Men{\'e}ndez-Delmestre}
  {et~al}\mbox{.}(2007){Men{\'e}ndez-Delmestre}, {Sheth}, {Schinnerer},
  {Jarrett}, \& {Scoville}}]{Menendez-Delmestre2007}
{Men{\'e}ndez-Delmestre} K., {Sheth} K., {Schinnerer} E., {Jarrett} T.~H.,
  {Scoville} N.~Z., 2007, \apj, 657, 790

\bibitem[{{Moles}, {Marquez} \& {Perez}(1995){Moles}, {Marquez}, \&
  {Perez}}]{Moles1995}
{Moles} M., {Marquez} I., {Perez} E., 1995, \apj, 438, 604

\bibitem[{{Mulchaey} \& {Regan}(1997)}]{Mulchaey1997}
{Mulchaey} J.~S., {Regan} M.~W., 1997, \apjl, 482, L135+

\bibitem[{{Pence}, {Taylor} \& {Atherton}(1990){Pence}, {Taylor}, \&
  {Atherton}}]{Pence1990}
{Pence} W.~D., {Taylor} K., {Atherton} P., 1990, \apj, 357, 415

\bibitem[{{Rix} \& {Zaritsky}(1995)}]{Rix1995}
{Rix} H.-W., {Zaritsky} D., 1995, \apj, 447, 82

\bibitem[{{Rots}(1975)}]{Rots1975}
{Rots} A.~H., 1975, \aap, 45, 43

\bibitem[{{Rots} {et~al}\mbox{.}(1990){Rots}, {Bosma}, {van der Hulst},
  {Athanassoula}, \& {Crane}}]{Rots1990}
{Rots} A.~H., {Bosma} A., {van der Hulst} J.~M., {Athanassoula} E., {Crane}
  P.~C., 1990, \aj, 100, 387

\bibitem[{{Rubin} {et~al}\mbox{.}(1985){Rubin}, {Burstein}, {Ford}, \&
  {Thonnard}}]{Rubin1985}
{Rubin} V.~C., {Burstein} D., {Ford}, Jr. W.~K., {Thonnard} N., 1985, \apj,
  289, 81

\bibitem[{{Rubin} {et~al}\mbox{.}(1982){Rubin}, {Ford}, {Thonnard}, \&
  {Burstein}}]{Rubin1982a}
{Rubin} V.~C., {Ford}, Jr. W.~K., {Thonnard} N., {Burstein} D., 1982, \apj,
  261, 439

\bibitem[{{Rubin}, {Ford} \& {.~Thonnard}(1980){Rubin}, {Ford}, \&
  {.~Thonnard}}]{Rubin1980}
{Rubin} V.~C., {Ford} W.~K.~J., {.~Thonnard} N., 1980, \apj, 238, 471

\bibitem[{{Ryder} {et~al}\mbox{.}(1998){Ryder}, {Zasov}, {Sil'chenko},
  {McIntyre}, \& {Walsh}}]{Ryder1998}
{Ryder} S.~D., {Zasov} A.~V., {Sil'chenko} O.~K., {McIntyre} V.~J., {Walsh} W.,
  1998, \mnras, 293, 411

\bibitem[{{Salo} {et~al}\mbox{.}(1999){Salo}, {Rautiainen}, {Buta}, {Purcell},
  {Cobb}, {Crocker}, \& {Laurikainen}}]{Salo1999}
{Salo} H., {Rautiainen} P., {Buta} R., {Purcell} G.~B., {Cobb} M.~L., {Crocker}
  D.~A., {Laurikainen} E., 1999, \aj, 117, 792

\bibitem[{{S{\'a}nchez-Gallego} {et~al}\mbox{.}(2012){S{\'a}nchez-Gallego},
  {Knapen}, {Wilson}, {Barmby}, {Azimlu}, \& {Courteau}}]{Sanchez-Gallego2012}
{S{\'a}nchez-Gallego} J.~R., {Knapen} J.~H., {Wilson} C.~D., {Barmby} P.,
  {Azimlu} M., {Courteau} S., 2012, \mnras, 2893

\bibitem[{{Schlegel}, {Finkbeiner} \& {Davis}(1998){Schlegel}, {Finkbeiner}, \&
  {Davis}}]{Schlegel1998}
{Schlegel} D.~J., {Finkbeiner} D.~P., {Davis} M., 1998, \apj, 500, 525

\bibitem[{{Schwarz}(1984)}]{Schwarz1984}
{Schwarz} M.~P., 1984, \mnras, 209, 93

\bibitem[{{Sellwood} \& {Wilkinson}(1993)}]{Sellwood1993}
{Sellwood} J.~A., {Wilkinson} A., 1993, Reports on Progress in Physics, 56, 173

\bibitem[{{Sheth} {et~al}\mbox{.}(2008){Sheth}, {Elmegreen}, {Elmegreen},
  {Capak}, {Abraham}, {Athanassoula}, {Ellis}, {Mobasher}, {Salvato},
  {Schinnerer}, {Scoville}, {Spalsbury}, {Strubbe}, {Carollo}, {Rich}, \&
  {West}}]{Sheth2008}
{Sheth} K. {et~al.}, 2008, \apj, 675, 1141

\bibitem[{{Sheth} {et~al}\mbox{.}(2010){Sheth}, {Regan}, {Hinz}, {Gil de Paz},
  {Men{\'e}ndez-Delmestre}, {Mu{\~n}oz-Mateos}, {Seibert}, {Kim},
  {Laurikainen}, {Salo}, {Gadotti}, {Laine}, {Mizusawa}, {Armus},
  {Athanassoula}, {Bosma}, {Buta}, {Capak}, {Jarrett}, {Elmegreen},
  {Elmegreen}, {Knapen}, {Koda}, {Helou}, {Ho}, {Madore}, {Masters},
  {Mobasher}, {Ogle}, {Peng}, {Schinnerer}, {Surace}, {Zaritsky},
  {Comer{\'o}n}, {de Swardt}, {Meidt}, {Kasliwal}, \& {Aravena}}]{Sheth2010}
---, 2010, \pasp, 122, 1397

\bibitem[{{Shlosman}, {Begelman} \& {Frank}(1990){Shlosman}, {Begelman}, \&
  {Frank}}]{Shlosman1990}
{Shlosman} I., {Begelman} M.~C., {Frank} J., 1990, \nat, 345, 679

\bibitem[{{Shlosman}, {Frank} \& {Begelman}(1989){Shlosman}, {Frank}, \&
  {Begelman}}]{Shlosman1989}
{Shlosman} I., {Frank} J., {Begelman} M.~C., 1989, \nat, 338, 45

\bibitem[{{Sicotte}, {Carignan} \& {Durand}(1996){Sicotte}, {Carignan}, \&
  {Durand}}]{Sicotte1996}
{Sicotte} V., {Carignan} C., {Durand} D., 1996, \aj, 112, 1423

\bibitem[{{Tully}(1974)}]{Tully1974}
{Tully} R.~B., 1974, \apjs, 27, 415

\bibitem[{{Tully} {et~al}\mbox{.}(2009){Tully}, {Rizzi}, {Shaya}, {Courtois},
  {Makarov}, \& {Jacobs}}]{Tully2009}
{Tully} R.~B., {Rizzi} L., {Shaya} E.~J., {Courtois} H.~M., {Makarov} D.~I.,
  {Jacobs} B.~A., 2009, \aj, 138, 323

\bibitem[{{van Albada}(1985)}]{vanAlbada1985}
{van Albada} G.~D., 1985, \aap, 142, 491

\bibitem[{{van Albada} \& {Roberts}(1981)}]{vanAlbada1981}
{van Albada} G.~D., {Roberts}, Jr. W.~W., 1981, \apj, 246, 740

\bibitem[{{van Albada}, {van Leer} \& {Roberts}(1982){van Albada}, {van Leer},
  \& {Roberts}}]{vanAlbada1982}
{van Albada} G.~D., {van Leer} B., {Roberts}, Jr. W.~W., 1982, \aap, 108, 76

\bibitem[{{van der Hulst}(1979)}]{vanderHulst1979}
{van der Hulst} J.~M., 1979, \aap, 75, 97

\bibitem[{{van der Hulst} {et~al}\mbox{.}(1992){van der Hulst}, {Terlouw},
  {Begeman}, {Zwitser}, \& {Roelfsema}}]{vanderHulst1992}
{van der Hulst} J.~M., {Terlouw} J.~P., {Begeman} K.~G., {Zwitser} W.,
  {Roelfsema} P.~R., 1992, in Astronomical Society of the Pacific Conference
  Series, Vol.~25, Astronomical Data Analysis Software and Systems I,
  {D.~M.~Worrall, C.~Biemesderfer, \& J.~Barnes}, ed., p. 131

\bibitem[{{Visser}(1980)}]{Visser1980}
{Visser} H.~C.~D., 1980, \aap, 88, 159

\bibitem[{{Walter} {et~al}\mbox{.}(2008){Walter}, {Brinks}, {de Blok},
  {Bigiel}, {Kennicutt}, {Thornley}, \& {Leroy}}]{Walter2008}
{Walter} F., {Brinks} E., {de Blok} W.~J.~G., {Bigiel} F., {Kennicutt}, Jr.
  R.~C., {Thornley} M.~D., {Leroy} A., 2008, \aj, 136, 2563

\end{thebibliography}

\bsp

%\label{lastpage}

\end{document}